\documentclass{ar-1col}
\usepackage[numbers,sort&compress]{natbib}
\jname{Ann. Rev. Cond. Mat. Phys.}
\jvol{10}
\jyear{2019}
\doi{10.1146/(doi tba)}

\usepackage{amsmath,amssymb}
\usepackage{bm}
\def\abs#1{{\lvert#1\rvert}} 
\def\svec#1{{\vec{#1}}} 
\def\rvec#1{{\bm#1}} 
\def\a2d{a_\text{2D}}
\def\lmf{{\ell_\text{mfp}}} 
\def\eb{{\varepsilon_B}} 
\def\ef{{\varepsilon_F}} 
\DeclareMathOperator{\Ei}{Ei} 
\let\Re\relax\DeclareMathOperator{\Re}{Re} 
\let\Im\relax\DeclareMathOperator{\Im}{Im} 

\begin{document}

\markboth{Enss and Thywissen}{Spin Transport of Unitary Fermions}
\title{Universal Spin Transport and Quantum Bounds for Unitary Fermions}
\author{Tilman Enss$^1$ and Joseph H. Thywissen$^2$ 
\affil{$^1$Institute for Theoretical Physics, University of Heidelberg, D-69120 Heidelberg, Germany; email: enss@thphys.uni-heidelberg.de}
\affil{$^2$Department of Physics, University of Toronto, Canada M5S 1A7; and Canadian Institute for Advanced Research, Toronto, Canada M5G 1M1; email: jht@physics.utoronto.ca}}

\begin{abstract}
We review recent advances in experimental and theoretical understanding of spin transport in strongly interacting Fermi gases. The central new phenomenon is the observation of a lower bound on the (bare) spin diffusivity in the strongly interacting regime. Transport bounds are of broad interest for the condensed matter community, with a conceptual similarity to observed bounds in shear viscosity and charge conductivity. We discuss the formalism of spin hydrodynamics, how dynamics are parameterized by transport coefficients, the effect of confinement, the role of scale invariance, the quasi-particle picture, and quantum critical transport. We conclude by highlighting open questions, such as precise theoretical bounds, relevance to other phases of matter, and extensions to lattice systems.
\end{abstract}

\maketitle

\tableofcontents

\section{INTRODUCTION}

Magnetization dynamics in liquids has been studied since the advent of nuclear magnetic resonance techniques. Spin echos were first observed in water, and in the presence of a field gradient were used to determine spin diffusivity \cite{Hahn:1950tv,Carr:1954wf,Torrey:1956tk}. In Fermi liquids, the same probes revealed spin-wave effects 
\cite{Silin:1958tc,Hone:1961fx,LR:1968,Leggett:1970,Corruccini:1971,Corruccini:1972vt,3HeWheatley,Meyerovich:1991,OwersBradley:1999uu}. Magnetization dynamics of non-degenerate gases were found to show closely related physics \cite{Bashkin:1981, Laloe:1982ww,Laloe:1982wc,Johnson:1984gr,Levy:1984du,Meyerovich:1985,Miyake:1985,Vasiliev:2012}.

The advent of ultracold atoms has created several new opportunities and renewed interest in spin dynamics. First, spin dynamics can be spatially resolved \cite{Mcguirk:2002,Du:2008de,Hulet2011,Koschorreck2013}. Second, the strength of the interactions can be tuned, enabling the study of strongly interacting systems \cite{Sommer:2011,Sommer:2011njp,Koschorreck2013,Bardon:2014,Trotzky:2015fe,Luciuk:2017iw} and dynamic control of interactions \cite{Du:2009hm}. Third, spin locking provides a pathway to improved coherent time for metrology applications \cite{Rosenbusch:2010,Krauser:2014ca}.

\begin{figure}[tb]
\includegraphics[width=5.5in]{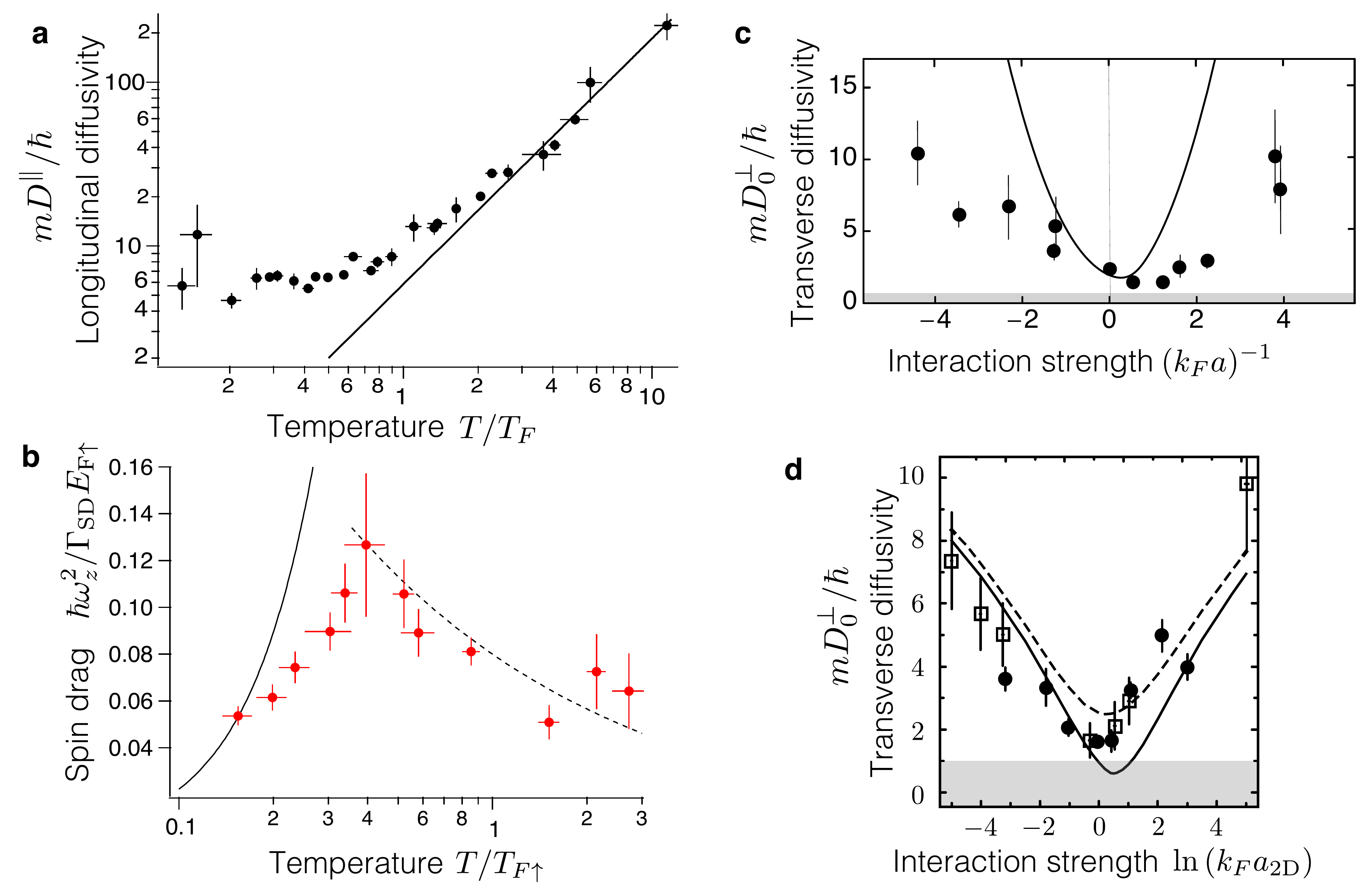}
\caption{\textbf{Evidence for dissipation bounds in spin dynamics.} 
\textbf{a} [from \cite{Sommer:2011}:] Longitudinal spin diffusivity versus reduced temperature $T/T_F$. At low temperature, the diffusivity approaches a constant value of $6.3(6) \hbar/m$ for a temperature below about 0.5$T_F$. Correcting for inhomogeneous gradients, the intrinsic value may be lower: see also  \textbf{Figure~\ref{fig:TrapTheory}a}.
\textbf{b} [from \cite{Sommer:2011njp}:] Normalized global relaxation time $\hbar\omega_z^2/\Gamma_\text{SD} E_{F\uparrow}$ of the spin dipole mode of the polarized Fermi gas as a function of the reduced temperature $T/{T_{F\uparrow}}$, where ${T_{F\uparrow}}$ is the local Fermi temperature at the center of the majority cloud. The solid curve is the low temperature limit $\propto T^2$ from~\cite{Bruun:2008}. The dashed curve is the high-temperature limit $0.08\sqrt{{{T_{F\uparrow}}}/{T}}$. 
\textbf{c} [from \cite{Trotzky:2015fe}:] Transverse spin diffusivity versus interaction strength for a 3D Fermi gas at initial $T/T_F=0.2$. Solid line shows a kinetic theory calculation for a uniform gas.
\textbf{d} [from \cite{Luciuk:2017iw}]: Transverse spin diffusivity versus interaction strength for a 2D Fermi gas with $T/T_F=0.31(2)$ (black circles) and $T/T_F=0.21(3)$ (open squares). The lines are predictions for $T/T_F=0.3$ by kinetic theory. The conjectured quantum bound would exclude the shaded areas in (c,d).
\label{fig:BoundedTransport} }
\end{figure}

This review focuses on ultracold fermionic systems in the near-unitary regime, where atoms become strongly interacting. The central new phenomenon is the observation of a lower bound in transport coefficients (\textbf{Figure~\ref{fig:BoundedTransport}}). In typical experiments, a high collision rate creates a fast local equilibration, which manifests as a slow global equilibration time. Bounds placed on local dissipation give {\em lower bounds} on transport coefficients, whereas bounds placed on system-wide equilibration are {\em upper bounds} on flow or communication \cite{Taylor:2015,Krinner:2017}. Transport equations connect the global to the local bounds, as discussed in \S\ref{sec:spintransport}. Bounded diffusivity, discussed in \S\ref{sec:TransCoeffs}, is found in the strongly correlated regime, where Fermi liquid theory breaks down. Experiments become a testing ground for powerful theoretical approaches such as holography, quantum critical dynamics, and numerical simulation.

A goal of this review is to connect spin dynamics in ultracold fermions to the broader discussion of dissipation, which also limits viscosity and conductivity. Study of quantum bounds provides physical insight into the nature of transport where perturbative calculations fail, and yet where universal limits relate diverse physical systems. Due to length constraints, we cannot cover all of vigorous activity in spin dynamics, so do not review lattice systems \cite{LewensteinLatticeReview}, integrable systems \cite{PolkovnikovRev}, spintronics \cite{spintronics}, or Kondo physics \cite{kondo}. We conclude in \S\ref{sec:conclusion} with a discussion highlighting some of the open questions in spin transport for strongly interacting fermions. 

\section{SPIN TRANSPORT \label{sec:spintransport}}

Spin transport describes how the local magnetization $\vec M(\rvec r,t)$ of a polarized system evolves in space and time. We start with the simplest hydrodynamic evolution equation and explain the basic phenomena of longitudinal and transverse spin transport. Microscopic physics is contained within transport coefficients, such as diffusivity and spin-rotation parameters, whose values for ultracold atomic gases we discuss in \S\ref{sec:TransCoeffs}.
\begin{marginnote}
\entry{$\vec{a}$}{Vector notation for spin-space, \emph{i.e.} $a_\alpha$ where $\alpha=\{x,y,z\}$.}
\entry{$\bm{a}$}{Spatial vector, \emph{i.e.} $a_i$ where $i=\{1,2,3\}$.}
\end{marginnote}

\subsection{Spin hydrodynamics \label{sec:spinhydro}}

The local magnetization obeys the continuity equation
\begin{align}
  \label{eq:magcont}
  \partial_t \vec M + \partial_i \vec J_i = \vec M \times \vec \omega_L.
\end{align}
The magnetization $\vec M(\rvec r,t)$ is a Bloch vector in spin space,
and it changes in time either by a spin current $\vec J_i(\rvec r,t)$
to a neighboring volume element in spatial direction $i$, or by
external torque with Larmor frequency $\vec\omega_L=\gamma_d \vec H_0$
in an external magnetic field $\vec H_0(\rvec r,t)$ with gyromagnetic
ratio $\gamma_d$.  In the frame rotating with $\vec \omega_L$, the
magnetization is conserved.  For a hydrodynamic fluid, the spin current is
given by 
\begin{align}
  \label{eq:spincur}
  \vec J_i = -D\bigl[\partial_i \vec M + \mu \vec M \times \partial_i 
  \vec M + \mu \vec M (\mu \vec M \cdot \partial_i \vec M) \bigr],
\end{align}
where $D = D_0/(1+\mu^2M^2)$ is an effective diffusion coefficient,
and $D_0$ is the bare diffusivity \cite{LR:1968,Leggett:1970}.
The Leggett-Rice parameter $\mu$ arises from exchange interactions and specifies
the strength of the spin-rotation effect (see \S\ref{sec:LR}). 
For $\mu=0$ one recovers Fick's law
$\vec J_i = -D_0 \partial_i \vec M$.  For general $\mu\neq0$ this is
replaced by the tensor equation
$\vec J_i = -\overleftrightarrow D \partial_i \vec M$ with
diffusivity tensor
$\overleftrightarrow D = D[\delta_{\alpha\beta} +
\varepsilon_{\alpha\beta\gamma} \mu M_\gamma + \mu^2 M_\alpha
M_\beta]$.
Although the evolution equation is nonlinear in $\vec M$, it can
be solved in experimentally relevant situations as shown below.

Even for isotropic $D_0$, the diffusivity tensor has two independent components with different diffusion coefficients in the low-temperature limit. The
gradient of the magnetization $\vec M=M\Hat{\svec e}$ can be decomposed into
a change of polarization magnitude $M$ and a change of spin direction $\Hat{\svec e}$,
\begin{align}
  \label{eq:gradmag}
  \partial_i \vec M = (\partial_i M) \Hat{\svec e} + M (\partial_i \Hat{\svec e}).
\end{align}
In the first term, $\partial_i \vec M$ is \emph{parallel} to $\vec
M$ and gives rise to a \emph{longitudinal spin current}
\begin{marginnote}
\entry{$\vec{M}$}{magnetization}
\entry{$\vec{J}_i$}{spin current in the $i$ spatial direction, often broken into $\vec J_i^\parallel$, $\vec J_i^\perp$}
\entry{$D_0$}{bare spin diffusivity}
\entry{$\mu$}{Leggett-Rice parameter}
\entry{$D$}{effective spin diffusivity}
\entry{$\overleftrightarrow D$}{diffusivity tensor, in Bloch space}
\end{marginnote}
\begin{align}
  \label{eq:spincurlong}
  \vec J_i^\parallel = -D_0 \partial_i \vec M
\end{align}
independent of $\mu$, with bare diffusivity $D_0$.  (This is seen from 
Eq.~\eqref{eq:spincur} where the second term vanishes and the third gives 
$\mu^2M^2\partial_i\vec M$.)
For a change of spin direction $\Hat{\svec e}$, the magnetization gradient
$\partial_i\vec M$ is \emph{perpendicular} to
$\vec M$ and gives rise to a \emph{transverse spin current}
\begin{align}
  \label{eq:spincurtrans}
  \vec J_i^\perp = -D [\partial_i \vec M + \mu \vec M
  \times \partial_i \vec M]
\end{align}
with reduced effective diffusivity $D\leq D_0$.
For $\mu\neq0$ the transverse current precesses around the local magnetization and exhibits the Leggett-Rice effect, discussed further in \S\ref{sec:LR}.

\subsubsection{Spin waves}
Spin waves have been studied extensively in both Bose and Fermi gases \cite{Mcguirk:2002, Oktel:2002, Fuchs:2002, Williams:2002, Fuchs:2003, Du:2008de, Du:2009hm, Piechon:2009, Natu:2009, Rosenbusch:2010, Koschorreck2013, Heinze:2013ko, Krauser:2014ca, Ebling:2014, Niroomand:2015fu, Xu:2015, Koller:2015, Xu:2017, Graham:2018}.  Transverse spin waves arise from the identical spin rotation effect (ISRE, see \S\ref{sec:LR} below).  Equivalently, a single spin-polarized atom traveling through a region of different spin polarization precesses and changes its internal spin state, in analogy to the Faraday effect.  For a trapped gas in the nondegenerate regime, the acquired phase accumulates over many passages through the gas and leads to anomalously large spin segregation \cite{Du:2008de, Du:2009hm, Piechon:2009, Natu:2009}.

\begin{figure}[tb!]
\includegraphics[width=4.5in]{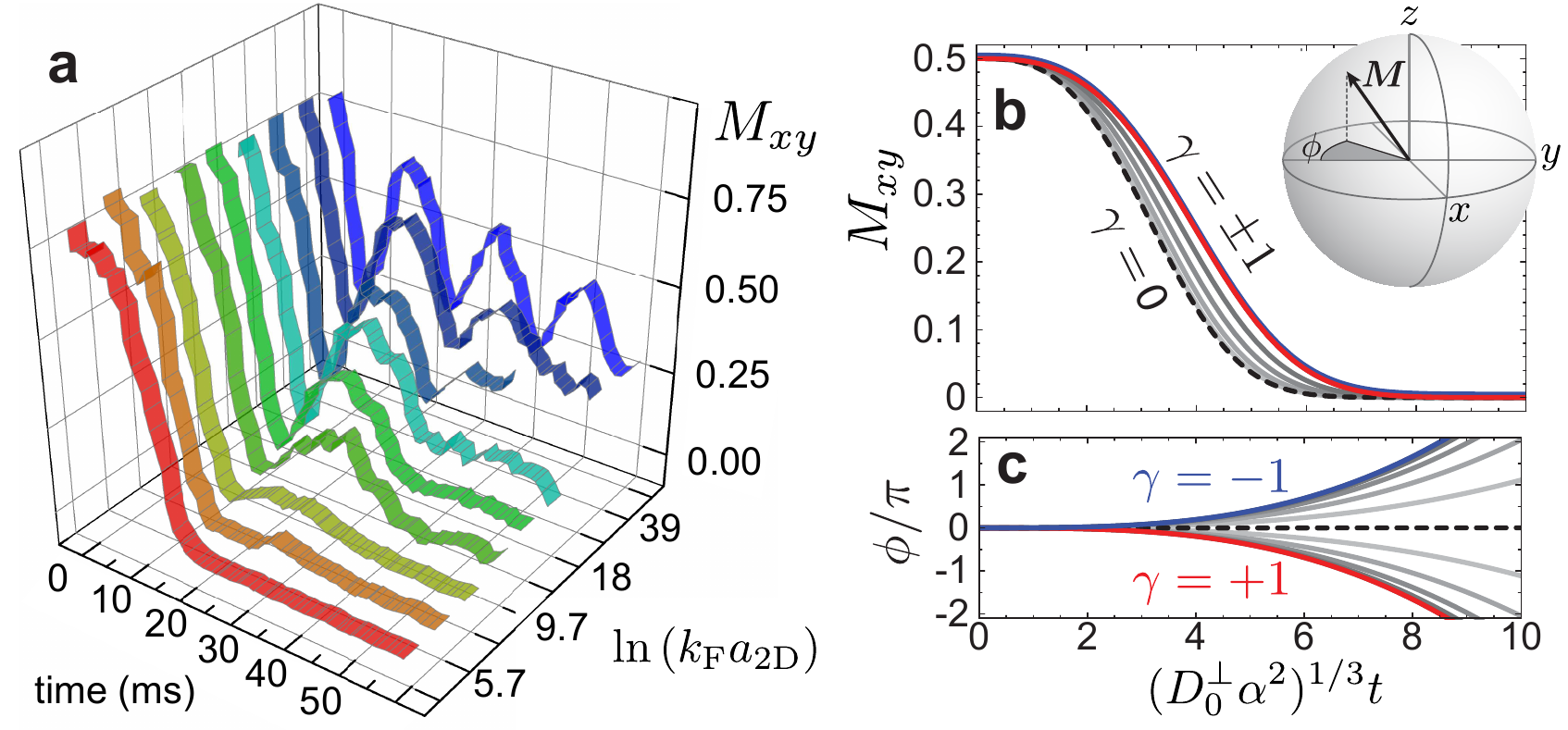}
\caption{\textbf{Transverse spin rotation.} Time evolution of transverse magnetization $M_{xy}$ observed through a Ramsey sequence. 
\textbf{a} [from \cite{Koschorreck2013}]: Under-damped transverse spin waves in a weakly interacting 2D Fermi gas (blue curves) are progressively more damped as interaction strength (inversely proportional to $\ln(k_F a_\text{2D})$) increases. 
\textbf{b,c} [from \cite{Trotzky:2015fe}]: In the over-damped regime, spin rotation modifies demagnetization dynamics. 
Here the case of tipping angle $\theta_1=5\pi/6$ and full initial polarization is plotted. Dashed lines in \textbf{b} and \textbf{c} show $\gamma=0$, and gray lines show steps of $0.2$ up to $\gamma=\pm 1$, where $\gamma$ is the dimensionless Leggett-Rice parameter, see Eq.~\eqref{eq:LRgamma}}
\label{fig:LRE}
\end{figure}

The continuity equation \eqref{eq:magcont} describes how the magnetization evolves for given spin current, and one can combine it with the constitutive relation \eqref{eq:spincur} for the spin current in terms of the magnetization to obtain a closed evolution equation for the magnetization:
$\partial_t \vec M = \partial_i \overleftrightarrow D \partial_i \vec M 
  + \vec M \times \vec \omega_L$.
In the rotating frame and for small deviations around equilibrium $\vec M = \vec M^\mathrm{(eq)} + \delta \vec M$, one has to first order in $\delta \vec M$
\begin{align}
  \label{eq:evolmag}
  \partial_t \, \delta \vec M
  = D[\delta_{\alpha\beta} + \varepsilon_{\alpha\beta\gamma}\mu M^\mathrm{(eq)}_\gamma + \mu^2 M^\mathrm{(eq)}_\alpha M^\mathrm{(eq)}_\beta] \nabla^2 \delta \vec M.
\end{align}
\begin{marginnote}
\entry{$\vec M^\mathrm{(eq)}$}{equilibrium magnetization}
\entry{$\delta \vec M$}{perturbation from $\vec M^\mathrm{(eq)}$}
\end{marginnote}
With a plane-wave ansatz $\delta \vec M \propto \exp{(i\rvec k\cdot\rvec r - i\omega t)}$ one finds one longitudinal and two transverse excitations in spin space, so-called \emph{spin waves} with frequencies
\begin{align}
  \label{eq:spinwavedisp}
  \omega_\parallel = -iD_0k^2 \quad \mbox{and} \quad 
  \omega_{\perp\pm} = -iD(1\pm i\mu M^\mathrm{(eq)})k^2 = -iDk^2 \pm \mu M^\mathrm{(eq)} Dk^2.
\end{align}
The longitudinal mode has frequency $\omega_\parallel$ independent of $\mu$; it is purely diffusive and has no dispersion (flat band).  On the other hand, there are two transverse modes $\omega_{\perp\pm}$ with complex effective diffusivity $\mathcal{D}_\text{eff} = D(1\pm i\mu M^\mathrm{(eq)})$. Its real part $D=D_0/(1+\mu^2 (M^\mathrm{(eq)})^2)$ depends on $\mu$ and is responsible for dissipation and attenuation of spin waves. The imaginary part of $\mathcal{D}_\text{eff}$, instead, gives rise to reactive response and a quadratic spin-wave dispersion with curvature $\mu M^\mathrm{(eq)}D$ in $k$ space.  Depending on the value of $\mu M^\mathrm{(eq)}$, there is a continuous crossover between the diffusive limit $\abs{\mu M^\mathrm{(eq)}}\gg1$ of nondispersing modes with real diffusivity governed by the diffusion equation, and the opposite limit $\abs{\mu M^\mathrm{(eq)}}\ll1$ of dispersing spin waves which are only weakly attenuated (see \textbf{Figure \ref{fig:LRE}a}). In the latter case, the effective diffusivity $\mathcal{D}_\text{eff}$ is nearly imaginary and describes the evolution under a Schr\"odinger equation with (nearly real) effective mass $m_\text{eff} = i/(2\mathcal{D}_\text{eff})$ \cite{Levy:1984du}.  This crossover is reminiscent of a damped harmonic oscillator between the overdamped and oscillating limits (see \S\ref{sec:trap} below).  For larger deviations $\delta \svec M$ from local equilibrium, longitudinal and transverse spin waves couple, see for instance \cite{Fomin:1997, Mullin:2005}.

\subsubsection{Leggett-Rice effect}
\label{sec:LR}
Spin rotation in the strongly diffusive regime was first analyzed by Leggett and Rice \cite{LR:1968,Leggett:1970}.
In the presence of an exchange term $\mu\neq0$, the spin current precesses around the local magnetization. This can be seen by re-writing \eqref{eq:spincur} as $\vec J_i = -D_0 \partial_i \vec M - \mu \vec J_i \times \vec M.$ 
The second term is reactive (\emph{i.e.}, nondissipative), rotating the spin current away from the direction of the magnetization imbalance and slowing magnetization decay. Both this slowing and the reactive term in the spin evolution are referred to as the ``Leggett Rice Effect'' (LRE). A two-body perspective on the same effect is the Identical Spin Rotation Effect \cite{Bashkin:1981, Laloe:1982ww, Laloe:1982wc}, in which exchange scattering in the collision term of a quantum gas rotates the spins out of their original plane in spin space. Note that this quantum effect in binary collisions remains visible even when the gas is not quantum degenerate. The perspectives have been shown to be equivalent in the weakly interacting regime \cite{Miyake:1985}.

The LRE is illustrated most clearly by considering a spin-echo sequence in an external magnetic field $\gamma_d \vec H_\text{ext} = (\omega_L + \alpha x_3)\Hat z$ along the spin $z$ direction, with a gradient of strength $\alpha = \gamma_d G$ along the spatial $x_3$ direction. The initial $z$ magnetization $\vec M=(0,0,M)$ is rotated by a $\theta_1$ pulse around the $y$ axis into $\vec M=(M\sin\theta_1,0,M\cos\theta_1)$ with finite transverse magnetization $M_x$.  Over time, the gradient winds the transverse magnetization into a spin helix in the $xy$ plane. After time $t/2$ a $\pi$ pulse is applied, after which the helix is unwound to give a spin echo at time $t$. Spin diffusion ruins the perfect re-alignment of spins and reduces the strength of the echo $\vec M(t)$ at the end of the sequence.
\begin{marginnote}
\entry{$\alpha$}{frequency gradient from magnetic field}
\entry{$\theta_1$}{rotation angle of initialization spin-flip}
\end{marginnote}

The transverse magnetization is conveniently parameterized by the complex number $\Tilde M_{xy} = M_x +iM_y$, and similarly for the spin current $J_{ixy} = J_{ix} + iJ_{iy}$.  The magnetization evolves with local Larmor frequency $\omega_L+\alpha x_3$ as
\begin{align}
  \label{eq:LRmagnplus}
  \partial_t \Tilde M_{xy} + \partial_i J_{ixy} = -i(\omega_L+\alpha x_3)\Tilde M_{xy} 
  \quad \mbox{and} \quad 
  \partial_t M_z + \partial_i J_{iz} = 0 \, .
\end{align}
For the given protocol only transverse spin currents arise,
\begin{align}
  \label{eq:LRspincur}
  J_{ixy}^\perp & = -D^\perp [(1+i\mu M_z) \partial_i \Tilde M_{xy} - i\mu \Tilde M_{xy} \partial_i M_z], \nonumber \\
  J_{iz}^\perp & = -D^\perp [\partial_i M_z + \mu \Im(\Tilde M_{xy}^* \partial_i \Tilde M_{xy})].
\end{align}
For homogeneous $D^\perp$ and $\mu$ the solution for the longitudinal magnetization $M_z(\rvec r,t)\equiv M_z$ remains constant in space and time \cite{Leggett:1970}.  We make an ansatz for a uniform magnetization in the rotating frame, $\Tilde M_{xy}(\rvec r,t) = e^{-i(\omega_L+\alpha x_3)t} M_{xy}(t)$, and are left with a simple time-evolution equation
\begin{marginnote}
\entry{$\tilde M_{xy}$}{$M_x + i M_y$}
\entry{$M_{xy}$}{$\tilde M_{xy}$ in the rotating frame}
\entry{$J_{i,xy}$}{$J_{i,x} + i J_{i,y}$}
\end{marginnote}
\begin{equation} \label{eq:LRmagn}
  \partial_t M_{xy} = -D^\perp (1+i\mu M_z) \alpha^2 t^2 M_{xy} \, .
\end{equation}
The transverse magnetization $M_{xy}(t) = \abs{M_{xy}(t)} e^{i\phi(t)}$ can be decomposed into real amplitude $\abs{M_{xy}(t)}$ and phase $\phi(t)$, which satisfy
\begin{align}
  \label{eq:LRampl}
  \partial_t \ln \abs{M_{xy}} = -D^\perp \alpha^2 t^2
  \qquad  \mbox{and} \qquad 
  \partial_t \phi = \mu M_z \partial_t \ln \abs{M_{xy}}\,.
\end{align}
The corresponding spin currents are
\begin{align}
  J_{3xy}^\perp  = D^\perp (1+i\mu M_z) i\alpha t \Tilde M_{xy}  \qquad  \mbox{and} \qquad 
  J_{3z}^\perp = D^\perp \mu \alpha t \abs{M_{xy}}^2 \, .
\end{align}
$J_{3xy}^\perp$ precesses around the local magnetization $M_z$ and accumulates a phase $\phi(t)$, while $J_{3z}^\perp$, which arises for $\mu\neq0$, is constant in space and has no observable effect.
In general, the effective diffusivity $D^\perp(t) = D_0^\perp / (1+\mu^2(\abs{M_{xy}(t)}^2+M_z^2))$ itself depends on time through the transverse magnetization.  For $\mu=0$ where $D^\perp = D_0^\perp$, or more generally for small tipping angle and transverse magnetization $\abs{M_{xy}}\ll M_z$, one finds the magnetization decay \cite{Leggett:1970}
\begin{align}
  \label{eq:LRapproxampl}
  \abs{M_{xy}(t)} & = \abs{M_{xy}(0)} e^{-D^\perp \alpha^2 t^3/3} = \abs{M_{xy}(0)} e^{-(R_M t)^3/3(1+\mu^2M_z^2)},\nonumber \\ 
  \phi(t) & = \mu M_z \ln(\abs{M_{xy}(t)} / \abs{M_{xy}(0)})
\end{align}
in terms of the bare diffusion rate $R_M = (D_0^\perp \alpha^2)^{1/3}$ which depends both on the diffusivity and the gradient strength.
\begin{marginnote}
\entry{$R_M$}{magnetization relaxation rate}
\end{marginnote}

In a spin-echo sequence, $t/2$ of winding and $t/2$ of un-winding yield a replacement $t^3/3 \mapsto t^3/12$ in the magnetization decay; whereas the phase $\phi(t)$ keeps accumulating since it is reversed synchronously with $M_z$. Equation \eqref{eq:LRampl} can be solved analytically \cite{Trotzky:2015fe} for arbitrary transverse magnetization. At the spin-echo time,
\begin{align}
  \label{eq:LRexactampl}
  \abs{M_{xy}(t)} & = \abs{M_{xy}(0)} \sqrt{ \eta^{-1} W(\eta \exp[\eta-2 D^\perp \alpha^2 t^3/3])}
\end{align}
with the Lambert $W$ function, the initial transverse amplitude $\eta \equiv \mu^2 \abs{M_{xy}(0)}^2 / (1+\mu^2 M_z^2)$, and the real part of the effective diffusivity $D^\perp = D_0^\perp / (1+\mu^2 M_z^2)$. The behavior is illustrated in \textbf{Figure~\ref{fig:LRE}b,c}. For small $\abs{M_{xy}(0)}$ it reproduces the Leggett solution \eqref{eq:LRapproxampl} 
but for $\eta\gtrsim1$ (large $\gamma=\mu n/2$ in the figure) the initial decay with $D^\perp(t=0) = D_0^\perp /(1+\eta)$ is much slower than the $\mu=0$ case and suggests an apparent diffusivity slowed down by a factor $1/(1+\eta)$, emphasizing the need to determine $\mu$ for accurate determination of $D^\perp_0$ from magnetization dynamics. For larger times, the magnetization decay deviates from the cubic exponential form and accelerates as $\abs{M_{xy}}$ itself decays.

\begin{figure}[tb!]
\includegraphics[width=5in]{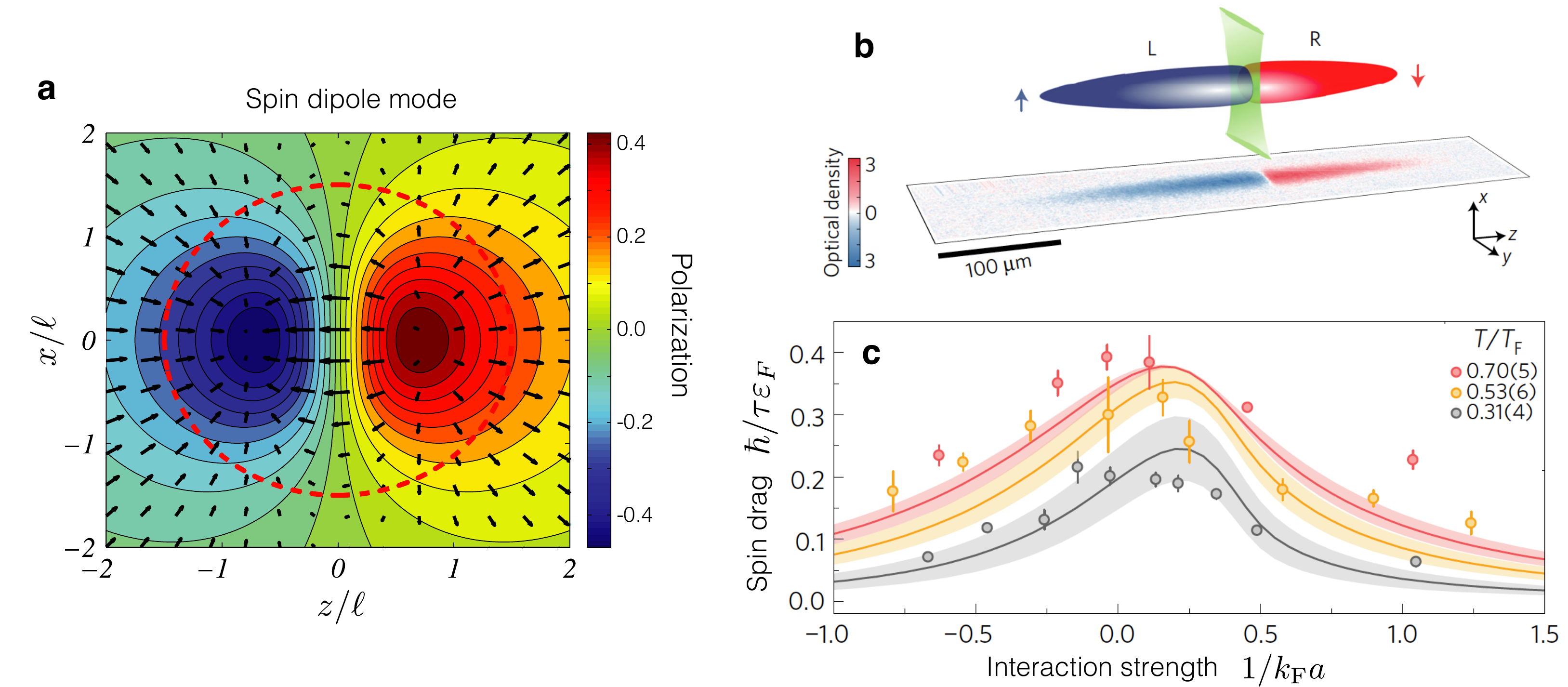}
\caption{\textbf{Spin dipole:} \textbf{a} [from \cite{BruunPeth:2011}]: Contour plots of the polarization and spin current density (arrows) in the $x z$-plane, for a longitudinal spin dipole mode excited along $z$. The red dashed contour shows where the density has fallen to $0.1$ of the central value. 
\textbf{b,c} [From \cite{Valtolina:2017}]: A Fermi gas is prepared by segregating the two spin components into two initially disconnected reservoirs at equilibrium by means of a thin optical barrier with a waist of about 2\,$\mu$m (green). Spin dipole dynamics show the \emph{local} spin relaxation rate $\hbar/\tau\ef$, plotted here as a function of $1/k_F a$ for  $0.31 \leq T/T_F \leq 0.7$.  Experimental points are obtained by fitting dynamics at $t>50\,$ms to the solution of the diffusion model \eqref{eq:spindipole} A maximum in the local spin relaxation rate $\tau^{-1}$ corresponds to a minimum in the global damping rate $\Gamma_\mathrm{SD}$. 
Lines are predictions from T-matrix kinetic theory, assuming the nominal initial $T/T_F$ and allowing a $\pm$20$\%$ temperature variation (shaded areas). These data and \textbf{Figure~\ref{fig:BoundedTransport}b} indicate that the minimum $\tau$ occurs at moderate degeneracy and near-resonant interactions.
\label{fig:SpinDipoleDamping}}
\end{figure}

\subsection{Longitudinal trap dynamics}
\label{sec:trap}
\begin{figure}[tb!]
\includegraphics[width=5.5in]{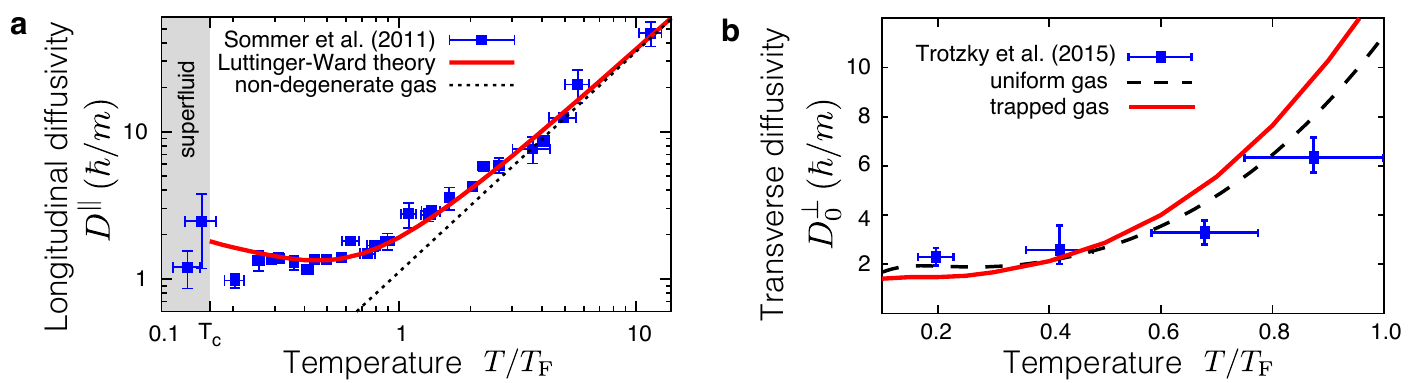}
\caption{\textbf{Inhomogeneity effects on spin transport} in a unitary Fermi gas. \textbf{a} [from \cite{Sommer:2011} and \cite{Enss:2012lw}:] Longitudinal diffusivity $D^\parallel$ vs.\ reduced temperature $T/T_F$ is compared to Luttinger-Ward calculation (solid red line) and the universal $T^{3/2}$ high-temperature behavior, $D^\parallel = 1.1\,(T/T_F)^{3/2} \hbar/m$ from kinetic theory \eqref{eq:D0_3D}. The experimental data \cite{Sommer:2011} (blue squares) for the trapped gas are rescaled down by a factor of $4.7$ to compensate for the effect of the trapping potential. At low temperature, the inferred local diffusivity approaches a constant value of $1.34(6) \hbar/m$. 
\textbf{b} [from \cite{Trotzky:2015fe} and \cite{Enss:2015er}:] Transverse diffusivity $D_0^\perp$ vs.\ reduced temperature $T/T_F$ is compared to uniform calculations (black dashed line) and calculations that include a trap (red solid line). The trap effects are relatively minor, compared to $D^\parallel$, near unitarity; but become significant when $\lmf$ becomes comparable to the system size.}
\label{fig:TrapTheory}
\end{figure}

Most experiments with ultracold atoms are performed in an external trapping potential, that is to first approximation harmonic.  The trapped gas has characteristic collective modes of density, where spin-up and -down move in phase, and spin, where they oscillate out-of-phase \cite{Vichi:1999}. In fact, the longitudinal component of the spin hydrodynamic equations depends strongly on local density \cite{BruunPeth:2011, Heiselberg:2012}; in contrast, the transverse component is insensitive to pure density gradients when the mean-free path $\lmf$ is shorter than the cloud size \cite{Enss:2015er}.  Therefore, even for a trapped gas, transverse spin transport in a strongly interacting Fermi gas probes essentially local properties (see \textbf{Figure \ref{fig:TrapTheory}b}). In the following, we focus on the more significant effects of a trap on longitudinal transport.

Collective spin dipole motion is observed experimentally by separating up- and down-spin clouds and then letting them collide, see \textbf{Figure~\ref{fig:SpinDipoleDamping}} \cite{Sommer:2011, Valtolina:2017}.  After initial bounces, the clouds merge through longitudinal spin diffusion.  This is exemplified by the spin-dipole mode shown in \textbf{Figure \ref{fig:SpinDipoleDamping}a}, which determines the motion of the relative position of the centers of mass $d(t)=\langle z_\uparrow(t)-z_\downarrow(t)\rangle$ of the spin-$\uparrow$ and $\downarrow$ clouds, and follows the equation of a damped harmonic oscillator:
\begin{align}
  \label{eq:spindipole}
  \ddot d(t) + \dot d(t)/\tau + \omega_z^2 d(t) = 0,
\end{align}
where $\omega_z$ is the trapping frequency and $\tau$ a characteristic \emph{local} relaxation time for longitudinal spin imbalance.  The complex solution $\omega = \sqrt{\omega_z^2-1/4\tau^2+i0}-i/2\tau = \omega_\text{SD} - i\Gamma_\text{SD}$ is characterized by the spin-dipole frequency $\omega_\text{SD}$ and its \emph{global} spin drag rate $\Gamma_\text{SD}$.  In the hydrodynamic limit $\omega_z\tau\ll1$ one has overdamped dynamics with $\omega_\text{SD} = 0$ and small drag rate $\Gamma_\text{SD} = \omega_z^2\tau$.  In the collisionless regime $\omega_z\tau\gg1$ one finds almost undamped oscillations $\omega_\text{SD} = \sqrt{\omega_z^2-1/4\tau^2}\to\omega_z$ with global drag rate $\Gamma_\text{SD} = 1/2\tau$ (see \textbf{Figure~\ref{fig:SpinDipoleDamping}c}).  Measurement of the spin-dipole mode gives thermodynamic information on the Fermi gas because $\omega_\text{SD}^2 = N/m\int d\rvec r\,z^2\chi$ is sensitive to the equation of state and in particular the trap integrated spin susceptibility $\chi(T,n,a)$ \cite{Recati:2011do}, or, for a polarized gas, the effective quasiparticle interaction \cite{Bruun:2008, Christensen:2015}.  Furthermore, the spin-drag rate $\Gamma_\text{SD}$ determines longitudinal spin diffusion at late times.
\begin{marginnote}
\entry{$d(t)$}{c.m.\ displacement of spin-$\uparrow$ and spin-$\downarrow$}
\entry{$\omega_z$}{trap frequency along excitation}
\entry{$\tau$}{local relaxation time}
\entry{$\omega_\mathrm{SD}$}{spin-dipole frequency}
\entry{$\Gamma_\mathrm{SD}$}{global damping rate}
\end{marginnote}

The collective trap dynamics are governed by hydrodynamic evolution in the dense trap center, but collisionless dynamics in the dilute outer regions of the trap.  It is a challenging task to extract local transport coefficients from global measurements of trap collective modes: an accurate determination requires not only knowledge of the local density profile and thermodynamic properties (via the local density approximation), but also of the local \emph{velocity} profile.  Inhomogeneous spin diffusion and spin drag have been computed at high temperature by an extension of the hydrodynamic diffusion equation to an inhomogeneous diffusion coefficient \cite{BruunPeth:2011, Taylor:2011} and by molecular dynamics simulation of the Boltzmann equation \cite{Goulko:2011, Goulko:2013}: at high $T$, fast diffusion occurs along the surface of the cloud and avoids the nondiffusive core.  In the quantum degenerate regime diffusion can only occur via the cloud center and is much slower \cite{Heiselberg:2012}.  With an appropriate density-dependent diffusivity, spin hydrodynamic equations can interpolate consistently between the dense and dilute regimes \cite{Schaefer:2016ga}.  In a more general case when both spin components have slightly different mass or experience slightly different trapping potentials, the center-of-mass mode decays very slowly while the different spin clouds experience almost perfect drag \cite{Bamler:2015}. For these reasons, the global minimum measured to be $\sim 6 \hbar/m$ in \textbf{Figure~\ref{fig:BoundedTransport}a} is interpreted to imply a local minimum close to $\sim \hbar/m$ at peak density (for instance, see \textbf{Figure~\ref{fig:TrapTheory}a}). 

\section{TRANSPORT COEFFICIENTS \label{sec:TransCoeffs} }

The spin transport coefficients $D_0$ and $\mu$ can be computed explicitly from microscopic models of ultracold quantum gases.  Specifically, one is interested in how the diffusivity $D_0(n,T,a)$ depends on spin component densities $n_{\uparrow/\downarrow}$, temperature $T$, and scattering length $a$.  For trapped systems, one often computes the transport properties first for a uniform system and then employs a local-density approximation to define a local diffusivity $D_0(\rvec r) = D_0\bigl(n(\rvec r),T(\rvec r),a(\rvec r)\bigr)$ as input for the hydrodynamic equations \eqref{eq:spincur} \cite{Sommer:2011, BruunPeth:2011, Enss:2015er, Schaefer:2016ga}.

First we argue that the values of transport coefficients for ultracold Fermi gases in fact determine universal scaling functions which apply to spin transport in generic short-range interacting Fermi gases.  Then we review spin transport measurements and calculations both for coherent quasiparticle transport and for incoherent transport near a quantum critical point.  Finally, we discuss universal bounds for transport coefficients.

\subsection{Universality and scaling}
\label{sec:universal}

Universality in dilute interacting 3D Fermi gases appears in different forms: for a short-range interaction $\abs{nr_e^3}\ll1$ the low-energy scattering is parametrized by the single $s$-wave scattering length $a$, and thermodynamic functions have a universal scaling form depending on the dimensionless interaction parameter $\lambda_T/a$, as long as all scales are larger than $\abs{r_e}$. 
However, the specific scaling form and in particular the scaling exponents depend on the region of the phase diagram: for small attractive $a<0$ the scaling functions are governed by the weakly coupled BCS fixed point of a dilute gas, while for strong binding with $a>0$ the Fermi gas falls into the universality class of a dilute Bose gas of molecules with weak repulsive interaction \cite{Sachdev:1999, Zwerger:2016}.  At both weak coupling fixed points, the correlation length diverges much faster than the particle spacing.%
\begin{marginnote}
\entry{$a$}{$s$-wave scattering length}
\entry{$r_e$}{Effective range of the interaction}
\entry{$\beta$}{$(k_B T)^{-1}$}
\entry{$\lambda_T$}{$\sqrt{2\pi\hbar^2\beta/m}$ is the thermal length}
\entry{$k_F$}{Fermi wave vector}
\end{marginnote}
Remarkably, at unitarity $a^{-1}=0$ there is a new, strongly interacting and nonperturbative fixed point, where the correlation length diverges along with the particle spacing as $n\to0$.  This unitary fixed point at zero temperature and density is a quantum critical point, which governs the whole phase diagram of the strongly interacting Fermi gas.  Near the fixed point, the scaling functions are not just generic functions of $\lambda_T/a$ but in addition their (anomalous) scaling dimensions are known, for instance the anomalous scaling of the three-body decay near unitarity \cite{Zwerger:2016}.

In the strongly interacting Fermi gas, both the thermodynamic and the dynamical (transport) properties exhibit universal scaling behavior of the form $f(T,\mu,a^{-1},h) = T^\alpha f_s(\beta\mu, \lambda_T/a, \beta h)$ in terms of a temperature scaling exponent $\alpha$ and a dimensionless scaling function $f_s$ which depends only on dimensionless ratios. 
\begin{marginnote}
\entry{$\mu$}{$(\mu_\uparrow + \mu_\downarrow)/2$ chemical potential, thermodynamic conjugate to particle number}
\entry{$h$}{$(\mu_\uparrow-\mu_\downarrow)/2$ differential chemical potential, thermodynamic conjugate to magnetization}
\end{marginnote}
In a canonical description at fixed total density $n=k_F^3/(3\pi^2)$ and Fermi energy $\ef=k_B T_F=\hbar^2k_F^2/2m$, the dimensionless scaling functions can also be expressed as $f_c(T/T_F,(k_F a)^{-1},h/\ef)$.

Once these universal scaling functions have been measured, or computed, for the strongly interacting ultracold Fermi gas, universality affords to make predictions about spin transport in diverse physical situations at vastly different energy scales, such as the strongly interacting fermionic neutron matter in the crust of neutron stars \cite{Gezerlis:2014, Strinati:2018}.  Beyond universality, the vicinity to the \emph{unitary} QCP further constrains the form of the scaling functions, for instance the absence of anomalous temperature scaling in the 3D shear viscosity $\eta = \hbar \lambda_T^{-3} f_\eta$.  Together with the entropy density $s = k_B \lambda_T^{-3}f_s$, this implies a \emph{finite} limit for the shear viscosity to entropy ratio $\eta/s = (\hbar/k_B) f_\eta/f_s$ as $T\to0$.

\subsection{Quasiparticle transport}

At high temperature $T\gg T_F$, kinetic theory and the Boltzmann equation for the atomic degrees of freedom (see \S\ref{sec:kinetic}) describe the transport processes as a function of $n,T,a$; this is justified in a virial expansion to a certain order in the fugacity as a small parameter.  A Fermi gas at low temperature, instead, is well described by Fermi liquid theory (see \S\ref{sec:FLT}) and the Landau-Boltzmann equation for long-lived fermionic quasiparticle degrees of freedom \cite{Baym:2008}.
In the normal state the quasiparticles correspond to dressed atomic states, while in the superfluid state they are Bogoliubov quasiparticles. In either limit, the kinetic paradigm relates the diffusivity in $d$ dimensions 
\begin{equation} \label{eq:diffkin}
  D_0 = v \lmf/d = {v^2}\tau_D/d
\end{equation}
to a typical quasiparticle velocity $v$ and to the mean free path $\lmf$, assuming that $\lmf$ is much larger than the particle spacing.
\begin{marginnote}
\entry{$\tau_D$}{Diffusive scattering time, or transport time}
\entry{$\lmf$}{$v \tau_D$, mean free path}
\entry{$\sigma_\mathrm{sc}$}{scattering cross-section}
\end{marginnote}
Hence, the diffusivity is large both in the high- and low-temperature limits, when spin is transported efficiently though weakly interacting quasiparticles with a large mean free path.

At intermediate temperatures near the superfluid-to-normal phase transition in a strongly interacting fluid, the mean free path extrapolates to the order of the particle spacing and the quasiparticle picture breaks down (see \S\ref{sec:qucrit}).  Transport is then dominated by incoherent relaxation at a rate given by the temperature, $k_B T/\hbar$, and the diffusivity attains minimum values of order $D_0 \simeq \hbar/m$ (see \S\ref{sec:bounds}).

\subsubsection{Kinetic theory at high temperature}
\label{sec:kinetic}

In the nondegenerate regime at high temperature, kinetic theory based on the Boltzmann equation describes spin transport and gives a unique value for the diffusivity in both the longitudinal and transverse channels, and for arbitrary polarization of the Fermi gas \cite{Jeon:1989}.  This is different from the Bose gas, where a difference in scattering lengths leads to anisotropy of the diffusivity in the nondegenerate regime \cite{Mullin:2006}.  In the dilute 3D Fermi gas the diffusive scattering time is \cite{Bruun:2011, Enss:2013}
\begin{equation} \label{eq:tau3D}
  \tau_D = \frac{3\pi\hbar\beta}{4\sqrt2n\lambda_T \sigma_\text{sc}},
  \qquad \mbox{where} \qquad
  \frac{\sigma_\text{sc}}{\lambda_T^2} = 1-x-x^2 e^x \Ei(-x) \leq 1
\end{equation}
and the scattering cross section $\sigma_\mathrm{sc}$ depends on the dimensionless ratio $x=\beta(\hbar^2/ma^2) = \lambda_T^2/2\pi a^2$ of the interaction energy scale to temperature, and $\Ei(-x)$ denotes the exponential integral.  Important limits are the unitary Fermi gas (UFG) with resonant interaction $\lambda_T/a\to0$ ($x=0$) and thermally limited cross section $\sigma_\mathrm{sc}=\lambda_T^2$ which yields $\tau_D^\text{UFG} = 3\pi\hbar\beta/4\sqrt2n\lambda_T^3$, as well as the weakly interacting limit $\abs a\to0$ with $\sigma_\mathrm{sc}=4\pi a^2$ and $\tau_D=\tau_D^\text{UFG} (\lambda_T^2/4\pi a^2)$.  Together with the thermal velocity $v^2/d=k_BT/m$ one obtains the high-temperature diffusion coefficient
\begin{align}
  \label{eq:D0_3D}
  D_0^\text{3D} & = \frac{k_BT\tau_D}m
  = \frac{3\pi\hbar}{4\sqrt2 m}\, \frac{1}{n\lambda_T^3}\, \frac{\lambda_T^2}{\sigma_\text{sc}}
  = \frac{9\pi^{3/2}\hbar}{32\sqrt2m}\, \left(\frac{T}{T_F}\right)^{3/2}\, \frac{\lambda_T^2}{\sigma_\text{sc}}.
\end{align}
The diffusivity has units $\hbar/m$ and scales inversely with the phase-space density $n\lambda_T^3 = (8/3\sqrt\pi)(T/T_F)^{-3/2}$.  At unitarity $D_0 \approx 1.1(\hbar/m)(T/T_F)^{3/2}$ grows as $T^{3/2}$ for high temperature at fixed density (see \textbf{Figure \ref{fig:BoundedTransport}a} and \textbf{\ref{fig:TrapTheory}a}). At weak coupling, $D_0 \approx 1.1(\hbar/m) (k_Fa)^{-2} (T/T_F)^{1/2}$.  Both $\tau_D$ and $D_0$ grow as $\sigma_\text{sc}$ decreases for weaker coupling or higher temperature; this is reflected in a smaller global damping rate $\Gamma_\text{SD} = 1/2\tau_D$ of trap collective modes \eqref{eq:spindipole} toward the collisionless limit.

The dimensionless Leggett-Rice parameter
\begin{align}
  \label{eq:LRgamma}
  \gamma=\mu n/2=-{\tau_D n W}/{\hbar}
\end{align}
is given in terms of the effective spin-spin interaction $W$,%
\begin{marginnote}
\entry{$\gamma$}{dimensionless Leggett-Rice parameter}
\end{marginnote}
which depends on the real part of the many-body scattering T matrix \cite{Jeon:1989, Enss:2013, Enss:2015er}. At high temperature, the T matrix $\mathcal T = -4\pi\hbar^2 f(k)/m$ is given by the two-particle scattering amplitude $f(k)=-1/(a^{-1}+ik)$.  In the unitary limit $f(k)=i/k$ is purely imaginary and $\gamma\propto\Re f=0$ vanishes without medium scattering, while at weak coupling $f(k)=-a$ yields $W=4\pi\hbar^2a/m$ and $\gamma=-3\lambda_T/(8\sqrt2a)$, which grows toward weaker coupling. The Leggett-Rice parameter is negative for repulsive interaction (and vice versa). 

In the 2D Fermi gas the diffusion time $\tau_D$ is given in terms of the effective 2D scattering cross section $\sigma_\text{sc}$ as
\begin{align}
  \label{eq:sigma2D}
  \tau_D = \frac{\pi\hbar\beta}{2n\lambda_T\sigma_\text{sc}}, \quad  
  \frac{\sigma_\mathrm{sc}}{\lambda_T} = \frac{\lambda_T^4}2 \int_0^\infty dk\,k^3 \frac{\exp(-k^2\lambda_T^2/2\pi)} {\ln^2(k^2\a2d^2)+\pi^2}
  \simeq \frac{\pi^2}{\ln^2(2\beta\eb/3)+\pi^2} \leq 1,
\end{align}
where $\eb=\hbar^2/m\a2d^2>0$ is the binding energy of the two-body bound state which is always present in the interacting 2D Fermi gas \cite{Bruun:2012, Enss:2012a}.  With the phase-space density $n\lambda_T^2=2T_F/T$ for the two-component Fermi gas at high temperature, the diffusivity is
\begin{align}
  \label{eq:D0_2D}
  D_0^\text{2D} & = \frac{k_B T\tau_D}{m}
  = \frac{\pi\hbar}{4m}\, \frac{T}{T_F}\, \left(1+\frac{\ln^2(2\eb/3k_B T)}{\pi^2}\right).
\end{align}
Here $D_0^\text{2D}$ is proportional to $\hbar/m$ and grows linearly with temperature, up to logarithmic corrections contained in the cross section $\sigma_\mathrm{sc}$. The diffusivity reaches a minimum value $D_0^\text{2D} \geq {\pi\hbar T}/({4m T_F})$ in the BEC-BCS crossover when $a \simeq \lambda_T$, and grows logarithmically toward weak coupling $\abs a\to0$.  The Leggett-Rice parameter  approaches $\gamma \simeq -\ln(2\eb/3k_B T)/\pi$ in the high-temperature limit.

Toward lower temperature quantum statistics start to play a role: in a fermionic scattering process the outgoing scattering states must be unoccupied, \emph{i.e.}, outside the Fermi surface.  This Pauli blocking reduces the scattering phase space at low temperature and leads to longer scattering times, and larger diffusivities than expected from the high-temperature limit \cite{Bruun:2011}.  At the same time, the medium scattering cross section for the remaining states is enhanced \cite{Enss:2013}, and the competition between smaller phase space and stronger medium scattering can even reduce the diffusivity below the classical value within kinetic theory, cf.\ {Figure~3} in Ref.~\cite{Enss:2012a}.  As one enters the quantum critical regime where the thermal length approaches the particle spacing, reliable results for this competition can be obtained from strong coupling approaches (see \S\ref{sec:qucrit}).

\subsubsection{Fermi liquid theory at low temperature}
\label{sec:FLT}

The study of spin transport in spin-polarized quantum systems has a long tradition in $^3$He, hydrogen, and other systems \cite{Silin:1958tc,Hone:1961fx,LR:1968,Leggett:1970,Corruccini:1971,Corruccini:1972vt,3HeWheatley,Meyerovich:1991}: the quantum statistics of identical particles and the noncommutativity of spin operators leads to two major effects, first a strong mean field which gives rise to spin waves and the Leggett-Rice effect, and second a giant increase of transport coefficients with polarization $P=M/n$ due to the Pauli principle.  For weakly interacting systems ($\abs{na^3}\ll1$) these effects are understood in kinetic theory at arbitrary temperature; in a degenerate polarized gas the difference of Fermi velocities of the majority and minority components leads to an anisotropy of the bare diffusion tensor in the direction along ($D_0^\parallel$) and perpendicular ($D_0^\perp$) to the magnetization \cite{Meyerovich:1985, Jeon:1989}.  However, nonlocal corrections appear already at the order of the mean field $\hbar\gamma P/k_BT\tau_D \sim (k_Fa)(T_F/T)P$.  Therefore, strongly interacting Fermi liquids ($\abs{k_Fa}\gtrsim1$) pose a challenge for Fermi liquid theory.  While longitudinal processes are still well described, transverse dynamics involve nonlocal interaction and imaginary off-shell scattering amplitudes beyond standard Fermi liquid theory.
\begin{marginnote}
\entry{$P$}{$M/n$ polarization of a spin-half Fermi system, between 0 and 1.}
\end{marginnote}

Let us first consider unpolarized Fermi liquids with isotropic diffusion, and then discuss the polarized case.  At low temperature $T\ll T_F$ in a normal Fermi liquid the fermionic quasiparticle excitations near the Fermi surface become long-lived and lead to a large transport time $\tau_D \propto T^{-2}$ and diffusivity \cite{Hone:1961fx, LR:1968, Leggett:1970}
\begin{marginnote}
\entry{Fermi Liquid parameters}{Mean-field response of an interacting Fermi gas}
\entry{$F_0^s$}{Scalar potential from a density fluctuation}
\entry{$F_1^s$}{Vector potential from a mass current}
\entry{$F_0^a$}{Effective field from a polarization}
\entry{$F_1^a$}{Spin vector potential from a spin current}
\end{marginnote}
\begin{align}
  \label{eq:DFL3D}
  D_0 = \frac{v_F^2}{3}(1+F_0^a)\tau_D \propto T^{-2}
\end{align}
in terms of the Landau parameters $F_\ell^\text{s,a}$ \cite{Baym:2008}.  Also the transport time can be expressed approximately in terms of Landau parameters; for the unitary Fermi gas these have been determined experimentally and yield $D_0 \approx 0.31 (\hbar/m) (T/T_F)^{-2}$ \cite{Bruun:2011}.  A generalization to the polarized unitary Fermi gas was given in \cite{Parish:2009}.

In two dimensions the diffusivity acquires additional logarithmic corrections, $D_0^\text{2D}\propto 1/[T^2\ln(1/T)]$ \cite{Fu:1974}.  The phase space for scattering in 2D is severely restricted such that only scattering angles $0$ and $\pi$ are allowed, hence the kinetic equation is exactly solvable and can be expressed in terms of Landau parameters without approximation \cite{Miyake:1984} -- an elegant example of an exact expression for diffusivity.

While in an unpolarized Fermi liquid the transverse spin diffusivity $D_0^\perp$ is identical to the longitudinal $D_0^\parallel$, in the polarized case they differ dramatically: instead of diverging as $D_0^\parallel \propto T^{-2}$ for low $T$, $D_0^\perp$ saturates toward a finite value as \cite{Meyerovich:1985, Jeon:1989, Meyerovich:1994}
\begin{align}
  \label{eq:Dperp}
  D_0^\perp \propto \frac1{T^2+T_a^2}
\end{align}
with an anisotropy temperature $T_a \sim M^2$ %
\begin{marginnote}
\entry{$T_a$}{anisotropy temperature, below which $D^\parallel$ and $D_0^\perp$ are distinct}
\end{marginnote}
that remains finite in the limit $T\to0$ and grows with polarization. This is due to the fact that all states between the majority and minority Fermi surfaces contribute to scattering at low energy, hence the scattering phase space and the transverse diffusive transport time $\tau_\perp$ remain finite in the zero temperature limit, and the diffusion coefficients read \cite{Meyerovich:1985}
\begin{equation}
  \label{eq:D0aniso}
  D_0^\parallel = \frac{\tau_\parallel}3\, \frac{nv_\uparrow^2v_\downarrow^2}{n_\uparrow v_\uparrow^2+n_\downarrow v_\downarrow^2} \qquad \mbox{and} \qquad
  D_0^\perp = \frac{\tau_\perp}5\, \frac{n_\uparrow v_\uparrow^2-n_\downarrow v_\downarrow^2}{n_\uparrow-n_\downarrow}
\end{equation}
in terms of the spin component densities $n_{\uparrow,\downarrow}$ and Fermi velocities $v_{\uparrow,\downarrow}$.  
This anisotropic diffusion leads to a modified constitutive relation replacing \eqref{eq:spincur} in the hydrodynamic Leggett-Rice equation,
\begin{align}
  \label{eq:LRaniso}
  \vec J_i & = -\frac{D_0^\perp}{1+\mu^2M^2} \bigl[(\partial_i \vec M)_T + \mu \vec M \times \partial_i \vec M \bigr] - D_0^\parallel (\partial_i \vec M)_L,
\end{align}
where the longitudinal projection of the magnetization gradient $(\partial_i \vec M)_L = \Hat M(\Hat M\cdot \partial_i \vec M)$ and the transverse $(\partial_i \vec M)_T = \partial_i \vec M - (\partial_i \vec M)_L$.  Equation \eqref{eq:LRaniso} applies both in the hydrodynamic and collisionless limits as long as $\abs{\mu M}\gg1$.  This modified LR equation has been implemented to simulate the magnetization dynamics of the trapped Fermi gas \cite{Enss:2015er}.  In a related effect, the attenuation of transverse spin waves in a polarized paramagnet ($D\sim k^2$) is much stronger than in the ferromagnetic Heisenberg model ($D \sim k^4$, see \cite{Forster:1975}) due to the presence of a spin current even at $T=0$ \cite{Mineev:2004}.

There has been a theory argument that long wavelength transverse spin currents were \emph{undamped} at $T=0$ \cite{Fomin:1997}, but inclusion of the collision term again introduces finite scattering \cite{Mullin:2005}, and in dilute helium mixtures a finite $D_\perp$ for low $T$ in equation~\eqref{eq:Dperp} was confirmed experimentally \cite{Akimoto:2003}.  For ultracold fermions at strong interaction, the temperature and polarization dependence of the transverse and longitudinal diffusivities has been computed within kinetic theory using the many-body T matrix which incorporates medium scattering \cite{Enss:2013, Enss:2015er}.  This makes it an interesting proposition to measure $D_0^\parallel$ and $D_0^\perp$ in ultracold Fermi gases, if one can reach low enough temperatures, and compare to theoretical predictions.

Finally, the Leggett-Rice parameter in a Fermi liquid is given by \cite{Leggett:1970}
\begin{align}
  \label{eq:FLgamma}
  \gamma & = -\frac{2n\tau_\perp}{\hbar N_F}\, \frac{F_1^a/3-F_0^a}{1+F_1^a/3} 
  = -2\lambda \frac{D_0^\perp}{\hbar/m^*}
\end{align}
in terms of the \emph{transverse} scattering time $\tau_\perp$, while the density of states $N_F$ yields $N_Fv_F^2/3 = n/m^*$.  The effective interaction $W$ in \eqref{eq:LRgamma} is captured by the dimensionless exchange interaction parameter
\begin{align}
  \label{eq:lambda}
  \lambda & = \frac{1}{1+F_0^a} - \frac{1}{1+F_1^a/3},
\end{align}
which is positive for repulsive and negative for attractive effective quasiparticle interaction.  At weak coupling to first order in $g=2k_Fa/\pi$, $F_0^a=-g$ and $F_1^a=0$ give $\lambda=1/(g^{-1}-1)$ which has the correct asymptotics but diverges at strong coupling $\abs g\geq1$ \cite{Miyake:1985, Meyerovich:1985}.  In order to experimentally determine $\lambda$ from Eq.~\eqref{eq:FLgamma}, one extracts the bare diffusivity $D_0^\perp$ from the magnetization decay \eqref{eq:LRexactampl} and the LR parameter $\gamma P=\mu M/2$ from the Ramsey phase $\phi$ in \eqref{eq:LRapproxampl} to find $\lambda$ from their ratio, $\lambda = -\hbar\gamma / 2m^*D_0^\perp$.  In the 3D unitary Fermi gas at low temperature a value of $\lambda \approx -0.2$ was measured, which yields the first experimental determination of the spin vector potential parameter $F_1^a \approx 0.5$ \cite{Trotzky:2015fe}; even larger values for $\lambda$ are found in 2D \cite{Luciuk:2017iw}.  Furthermore, the sign change of $\lambda$ reveals at which point in the BEC-BCS crossover the effective interaction turns from repulsive to attractive, because the direction of spin rotation depends on an effective spin-spin interaction (see Eq.~\ref{eq:lambda}).

\subsubsection{Symmetry broken phases}
At low temperature the \emph{attractive} Fermi gas enters a superfluid (SF) phase. Mesoscopic transport experiments with near-unitary ultracold SF have observed suppression of spin conductance, consistent with a spin insulator \cite{Krinner:2016,PhysRevLett.117.255302,Zhang:2017,Ueda:2017}. The low-energy phonon excitations of the superfluid are responsible for particle and mass transport, while spin transport occurs exclusively via thermal fermionic excitations.  Here, we shall follow the derivation by Einzel \cite{Einzel:1991}. The fermionic excitations are Bogoliubov quasiparticles with dispersion $E_k=\sqrt{\xi_k^2+\Delta^2(T)}$ and isotropic gap parameter $\Delta(T)$ for $s$-wave pairing.  Assuming that the external magnetic field is small, $\omega_L\ll\Delta$, and does not destroy superfluidity, the Bogoliubov quasiparticles follow a Fermi distribution.  The thermal occupation number is exponentially suppressed at low temperature, $n_\text{qp}(T) = 2\sum_k f(E_k) \overset{T\to0}{\longrightarrow} 2\ln2 N_F k_B T Y_0(T)$ in terms of the Yosida function $Y_0(T) = -\int_{-\infty}^\infty d\xi_k\,(\partial f(E_k)/\partial E_k) \overset{T\to0}{\longrightarrow} \sqrt{2\pi\Delta/k_B T} \exp(-\Delta/k_B T)$, where $f(E)$ is the Fermi distribution and $N_F$ the density of states at the Fermi surface.  Importantly, the Bogoliubov quasiparticle lifetime from 2-body collisions is exponentially enhanced by the same factor, $\tau_D(T) \propto 1/Y_0(T)$, and hence the spin conductivity $\sigma_s$ has a \emph{finite} limit for low $T$.  In the superfluid phase the magnetic susceptibility $\chi(T) = \chi_0(T)/[1+F_0^a\chi_0(T)]$ with $\chi_0(T)=Y_0(T)$ is again exponentially small, and hence the spin diffusivity \cite{Einzel:1991}
\begin{align}
  \label{eq:DSF3D}
  D_0^\text{SF} & = \frac{v_\text{rms}^2}{3} [1+F_0^a Y_0(T)] \tau_D(T) \propto \exp(\Delta/k_B T)
\end{align}
is finite at $T_c$ and grows exponentially for $T\to0$.  In contrast, a pseudogap ansatz finds the spin conductivity exponentially suppressed at low $T$, while the diffusivity decreases moderately below $T_c$ \cite{Wulin:2011,Mueller:pseudogaps}. 
The Leggett-Rice effect applies also to Bogoliubov quasiparticles: for $s$-wave pairing, isotropic bare diffusivity gives rise to anisotropic effective diffusion in presence of the exchange interaction.  In contrast, in the A phase of $^3$He with $p$-wave pairing the Leggett-Rice effect is present already by the tensor structure of the bare diffusivity even without exchange interaction \cite{Einzel:1991}.

On the other hand, a sufficiently \emph{repulsive} Fermi liquid at low temperature exhibits an instability toward a ferromagnetic (FM) state, which is again a spin insulator.  Just above $T_c$, the spin drag rate $1/\tau_D(T) - 1/\tau_D(T_c) \propto (T-T_c)\ln(T-T_c)$ becomes small as $T\searrow T_c\ll T_F$ approaches the critical temperature from above \cite{Duine:2010}.  With $\tau_D(T_c)$ finite, the spin susceptibility $\chi(T) \propto 1/(T-T_c)$ determines the spin diffusivity
\begin{align}
  \label{eq:DFM3D}
  D_0^\text{FM}(T) = n\tau_D/m\chi \propto (T-T_c)^\kappa,
\end{align}
which vanishes near $T_c$ with exponent $\kappa=1$ at mean-field level.  Close to the transition, critical fluctuations in the universality class of the Heisenberg ferromagnet give $\kappa = (1-\eta)\nu/2$ in terms of the static critical exponents $\eta$ and $\nu$.

Inside the FM state ($F_0^a<-1$), the sample is spontaneously magnetized without external field.  In this regime, low-energy excitations are dispersing spin waves with a \emph{negative} damping rate $\Re D_0^\text{FM} \propto -\abs{M(T)}^2$ \cite{Mineev:2005js}.  Spin waves are thus found to be dynamically unstable toward transversal inhomogeneous magnetization perturbations, which might render a pure Fermi liquid description inapplicable.  With ultracold fermions this regime can also be approached from the fully polarized limit by studying the transport and scattering properties of a single mobile impurity, the Fermi polaron \cite{Bruun:2008, Christensen:2015, Valtolina:2017}.

\subsection{Quantum critical transport}
\label{sec:qucrit}

In the strongly interacting 3D Fermi gas, scaling is governed by a quantum critical point (QCP) at $T=0$, $\mu=0$, $a^{-1}=0$ and $h=0$ \cite{Sachdev:1999, Nikolic:2007, Zwerger:2016} introduced in \S\ref{sec:universal}. 
Deviations from the QCP to finite values of $T$, $\mu$, $a^{-1}$ and $h$ are relevant perturbations, which change the properties of the Fermi gas according to critical scaling functions $f(T,\mu,a^{-1},h)$.  Universality means that $f$ depends only on these relevant perturbations and no other system parameters in a whole neighborhood of the QCP, which in the case of the dilute Fermi gas extends up to the microscopic van-der-Waals scale $E_\text{vdw} = \hbar^2/m\ell_\text{vdw}^2$ \cite{Zwerger:2016}.  
\begin{marginnote}
\entry{Quantum Critical Point}{Conditions in which a phase transition occurs at zero temperature.}
\end{marginnote}

The unitary Fermi gas has a low-temperature superfluid and a high-temperature normal Fermi liquid phase.  In the $T$-$\mu$ phase diagram at unitarity $a^{-1}=h=0$, the superfluid phase boundary $\beta\mu_c\approx2.5$ \cite{Ku:2012, Enss:2012crit} separates the superfluid state $\beta\mu>\beta\mu_c$ from the normal state for $\beta\mu<\beta\mu_c$.  The normal state crosses over from a dilute classical gas at $\beta\mu\lesssim -1$ to a quantum degenerate gas at $\beta\mu\gtrsim -1$.  In the quantum critical region $-1\lesssim\beta\mu<\beta\mu_c$ above the QCP, many low-energy critical fluctuations are excited thermally and dominate the physical response and the scaling functions.  In this region the thermal length $\lambda_T$ is comparable to the particle spacing $n^{-1/3}$, hence thermal and quantum effects are equally important \cite{Sachdev:1999}.  Due to abundant quantum critical fluctuations, a local perturbation relaxes generically at a rate $\tau^{-1} \lesssim k_B T/\hbar$ determined only by temperature and the Planck constant and not by microscopic system parameters, with a \emph{universal} coefficient of order unity \cite{Sachdev:1999,Zaanen:2004he}. This applies to relaxation of any nonconserved quantity such as heat, shear stress, or spin current (however, see \cite{Farinas:1999} for conserved spin current in 2D).  Instead, the conserved particle number or momentum current are governed by the much slower hydrodynamic evolution.  In particular, for the nonconserved spin currents the transport scattering times can be expressed as universal functions
\begin{marginnote}
\entry{planckian dissipation}{Relaxation on the fast time scale $\hbar/k_B T$}
\end{marginnote}
\begin{align}
  \label{eq:tauQC}
  \tau_{\parallel,\perp} = \frac{\hbar}{k_B T} f_{\parallel,\perp}(\beta\mu,\lambda_T/a,\beta h),
\end{align}
and similarly for the dimensionless Leggett-Rice parameter $\gamma = f_\gamma(\beta\mu,\lambda_T/a,\beta h)$.  

The Einstein relation gives the longitudinal diffusivity $D_\parallel = \sigma_\parallel / \chi_\parallel$ as a ratio of spin conductivity $\sigma_\parallel = (1/\hbar \lambda_T) f_\sigma(\beta\mu,\lambda_T/a)$ and static spin susceptibility $\chi_\parallel = (\beta/V)\langle (N_\uparrow - N_\downarrow)^2 \rangle = \beta\lambda_T^{-3} f_\chi(\beta\mu,\lambda_T/a)$ in terms of dimensionless scaling functions, such that
\begin{marginnote}
\entry{Spin conductivity}{$\sigma$}
\entry{Spin susceptibility}{$\chi$}
\entry{Einstein relation}{$D=\sigma/\chi$}
\end{marginnote}
\begin{align}
  \label{eq:Dscaling}
  D_\parallel = \frac{\sigma_\parallel}{\chi_\parallel}
  = \frac{2\pi\hbar}{m} \, \frac{f_\sigma}{f_\chi}.
\end{align}
The dynamical longitudinal spin-current correlation function \cite{Enss:2012lw}
\begin{align}
  \label{eq:spincorr}
  g_\sigma(t) & = i\int d\rvec x \left\langle [j_\uparrow^z-j_\downarrow^z(\rvec x,t), j_\uparrow^z-j_\downarrow^z(\rvec 0,0)]\right\rangle
\end{align}
becomes a universal scaling function $g_\sigma(\beta\mu,\lambda_T/a,t) = (1/\hbar^2\beta^2\lambda_T) \Phi_\sigma(\beta\mu,\lambda_T/a,t/\hbar\beta)$ near the quantum critical point, where time is expressed in units of the thermal time scale $\hbar\beta$.  The derivation of transport bounds below assumes that $g_\sigma$ is finite in the zero-range limit, \emph{i.e.}, that the scaling function has no anomalous dimension, as found by explicit calculation in the zero-range model \cite{Enss:2012lw}.  Although the frequency dependent spin conductivity of the zero-range model has an anomalously slow decay for large frequency as $\omega^{-3/2}$ in 3D \cite{Hofmann:2011, Enss:2012lw}, it satisfies the usual spin f-sum rule $\int d\omega\, \sigma_s(\omega)/\pi = n/m$ \cite{Enss:2013sumrule}.
Then by the Kubo formula, the longitudinal spin conductivity is given by \cite{Zwerger:2016}
\begin{align}
  \label{eq:spincond}
  \sigma_\parallel(\beta\mu,\lambda_T/a)
  & = \frac1{\hbar} \int_0^\infty dt\, t\, g_\sigma(\beta\mu,\lambda_T/a,t) 
  = \frac{f_\sigma}{\hbar\lambda_T}, &
  f_\sigma & = \int_0^\infty dy\,y\Phi_\sigma(\beta\mu,\lambda_T/a,y)
\end{align}
in terms of rescaled time $y=t/\hbar\beta$.  

Explicit computations of these scaling functions in the quantum critical region are complicated by the absence of a small expansion parameter.  Results for $D_\parallel$ have been reported from a pseudogap ansatz \cite{Wulin:2011}, from a self-consistent conserving Luttinger-Ward approach \cite{Enss:2012lw}, by quantum Monte Carlo \cite{Wlazlowski:2013}, and by expanding in artificial small parameters at strong coupling, such as a $1/N_f$ expansion in the number of fermion flavors $N_f$ \cite{Enss:2012crit}.  Comparison with experiment is reported in \textbf{Figure~\ref{fig:TrapTheory}a}.

When approaching a SF or FM phase transition from the normal state, critical fluctuations lead at the same time to enhanced single-particle scattering and reduced single-particle density of states.  Their effect is to increase the spin relaxation rate $1/\tau_\parallel$, and reduce the diffusivity $D_\parallel$ approaching the transition \cite{Duine:2010, Mink:2012, Mink:2013, Wlazlowski:2013}.

\subsection{Transport bounds}
\label{sec:bounds}

As shown above, the spin diffusivity $D_0^\parallel$ becomes large both in a nondegenerate gas at high temperature (Eq.~\eqref{eq:D0_3D}) and at low temperature (Eq.~\eqref{eq:DSF3D}, see also discussion in \cite{Einzel:1991,Mineev:2005js,Zwerger:2016}). In between, there must be a minimum value where spin diffusion is slowest. Experimental results support this: for the 3D unitary Fermi gas, a minimum value $D_0^\parallel \simeq 1.3\hbar/m$ is observed in the quantum degenerate regime \cite{Sommer:2011} at $T/T_F\simeq 0.5$; similarly, in the transverse channel $D_0^\perp \simeq 2.3\hbar/m$ is observed \cite{Trotzky:2015fe} at $T/T_F\simeq 0.2$ (see \textbf{Figure~\ref{fig:TrapTheory}}). This is an example of a transport bound where quantum mechanical scattering imposes a lower bound on a transport coefficient, setting a quantum limit of diffusion in the absence of any scale in the problem apart from $\hbar$ and the particle mass $m$. 

Such a bound is reminiscent of the conjectured lower bound for the ratio of shear viscosity and entropy density, $\eta/s\geq \hbar/(4\pi k_B)$, which corresponds to minimal friction or ``perfect fluidity'', the closest any real fluid can come to being an ideal fluid \cite{Schaefer:2009}.  In fact, several strongly interacting quantum fluids come close to this bound, and the unitary Fermi gas has the lowest viscosity $\eta/s\gtrsim 0.5\hbar/k_B$ of any nonrelativistic fluid found so far, corresponding to a shear diffusion rate $D_\eta = \eta/mn \gtrsim 0.5\hbar/m$ \cite{Enss:2011, Zwerger:2016}.

Within kinetic theory, the diffusivity is $D_0 =v\lmf/d$, where $v\simeq v_F=\hbar k_F/m$ in the quantum degenerate regime.  The Mott-Ioffe-Regel (MIR) limit stipulates that the mean free path $\lmf\gtrsim 1/k_F$ is bounded from below by the particle spacing in the absence of any other length scale, which therefore could be saturated at strong coupling in the unitary regime where $a^{-1}\to0$.  This argument would suggest a quantum bound $D_0 \gtrsim \hbar/m$ \cite{Sommer:2011}, as found above.  Kinetic theory, however, assumes the existence of well defined quasiparticles, which is not guaranteed near the superfluid transition where the minimum is reached.  Instead, transport minima appear in the quantum critical regime above a zero-density quantum critical point \S\ref{sec:qucrit}.  They follow from the universality of the scaling functions and amplitude ratios and make no reference to the mean free path, which is not well defined in this regime \cite{Zwerger:2016}.  

The minimum $D_\parallel\gtrsim 1.3\hbar/m$ is reached near $T_c$ in the quantum critical regime at $\beta\mu\simeq0.3$ \cite{Enss:2012lw, Zwerger:2016}, see \textbf{Figure~\ref{fig:TrapTheory}a}.  In this region, the spin conductivity and susceptibility are only known numerically from Luttinger-Ward \cite{Enss:2012lw}, pseudogap \cite{Wulin:2011} and quantum Monte Carlo computations \cite{Wlazlowski:2013, Jensen:2018}.  While the spin susceptibility in the unitary Fermi gas varies only slowly with temperature $\chi_\parallel \simeq 0.4\chi_0$ above $T_c$ for $T_c < T \lesssim 0.5T_F$ in terms of the susceptibility $\chi_0=3n/2\ef$ of the ideal Fermi gas at zero temperature \cite{Enss:2012lw, Jensen:2018}, it drops steeply below $T_c$.  Equivalently, the spin conductivity exhibits a minimum $\sigma_\parallel \gtrsim 0.8\hbar n/m\ef$ above $T_c$ \cite{Enss:2012lw}, which corresponds to a maximum in the local spin drag rate $1/\tau_\parallel = n/m\sigma_\parallel$ where $D_\parallel = (v_F^2/3) (\chi_0 \tau_\parallel/\chi_\parallel)$.  Instead, other QMC and pseudogap calculations find that $\chi_\parallel$ is suppressed already above $T_c$ for $T\lesssim0.25T_F$ by a pseudogap effect \cite{Wulin:2011, Wlazlowski:2013}; the spin drag rate $\tau_\parallel^{-1}$ then grows toward lower temperatures also below $T_c$ and leads to a spin diffusivity $D_\parallel \simeq 0.8\hbar/m$ near $T\simeq0.1T_F$ \cite{Wlazlowski:2013}.  Measurements of longitudinal diffusion in the SF phase might resolve this question.

For the homogeneous unitary Fermi gas in the quantum critical regime, the particle spacing is the only independent length scale in the system (the thermal length is comparable and at most a factor of three larger in the temperature range $T_c < T \lesssim T_F$ above a high-temperature superfluid with $T_c\simeq 0.16T_F$ \cite{Ku:2012}).  
Hence, the diffusivity minimum $D\gtrsim \hbar/m$ together with a maximum susceptibility $\chi\lesssim n/\ef$ corresponds to a maximum drag rate $\tau_\parallel^{-1} \lesssim \ef/\hbar$ and a minimum transport time $\tau_\parallel \gtrsim \hbar/\ef$ and conductivity $\sigma_\parallel \gtrsim \hbar n/m\ef$ (all these bounds are equivalently expressed in terms of thermal length $\lambda_T$ and $\beta$).  In this strongly interacting regime there is no separation of scales which would allow expansions in a small parameter, and no proper quasiparticles exist.  Indeed, the absence of well defined quasiparticles appears to be a prerequisite for having transport bounds because if all excitations were long-lived, the transport time $\tau$ would also be large.  For longitudinal diffusion a transport minimum occurs near $T_c$ because both the high-$T$ nondegenerate and the low-$T$ superfluid limits have well defined quasiparticles.  This is remarkably different for transverse diffusion, which exhibits finite scattering down to zero temperature, and might therefore diffuse very slowly in the entire range $T\ll T_F$.

Quantum-limited (or ``planckian'') relaxation rates are also inferred from conductivity of strongly correlated electronic systems and heavy-fermion compounds in regimes of $T$-linear resistivity \cite{Bruin:2013}.
For electrical transport in metals, one can distinguish coherent and incoherent transport \cite{Hartnoll:2015}: for coherent transport the electrical current is almost conserved and its slow decay is governed by non-universal, material-specific properties such as lattice structure and disorder; in this case no universal bounds are expected. Incoherent transport arises when no quasiparticles exist due to strong electron interaction and the current decays fast: then its decay rate is governed by universal properties of strongly interacting quantum systems, and one obtains universal bounds. 
Spin transport in a strongly interacting Fermi gas clearly falls into the second category because there are no good quasiparticles, and \emph{spin} current is not conserved since it decays generically by interaction. Spin transport in ultracold fermions can therefore explore \emph{universal} incoherent spin conductors and their diffusion bounds. 

Finally, we note that an upper bound on local diffusivity $D\lesssim v^2\tau_\text{eq}$ is obtained when demanding that diffusive motion after local equilibration, $t\gtrsim \tau_\text{eq}$, remains inside the Lieb-Robinson light cone $\abs x<vt$ of entanglement spreading \cite{Hartman:2017}.

\section{DISCUSSION AND CONCLUSION \label{sec:conclusion} }

We conclude with a discussion of the broader significance of spin transport bounds and open questions raised from the theory and observations of spin transport.

Transport bounds have been studied primarily without optical lattices in ultracold atoms. \textbf{Bridging between these studies and materials would be further studies of spin transport \cite{Fukuhara:2013,Hild:2014it,Nichols:2018} and charge transport \cite{anderson2017optical,brown2018bad} of strongly interacting fermions in optical lattices.} 
A lattice breaks the scale invariance of the unitary Fermi gas. In some limits, transport bounds may not apply: for instance, applying extrinsic disorder potential can induce a localized phase in which diffusivity would have no lower bound. Similarly, low-temperature phases of lattices are insulating, and known to violate metallic conductivity bounds. From studies in materials, it has long been known that the MIR limit can be violated if scattering by other degrees of freedom introduces new length scales (effective range, lattice constant, localization length, etc.) and separate conditions for unitary scattering.  For example, in bad metals \cite{Emery:1995prl} the transport scattering time and the diffusivity saturate at high temperature when the mean free path reaches the lattice constant, but the resistivity grows linear in $T$ beyond the MIR limit by decreasing carrier density \cite{Perepelitsky:2016}. By comparison, spin conductivity of fermions in an optical lattice $\sigma \sim t/U\hbar$ parametrically violates the MIR limit $\sigma_0\simeq 1/\hbar$ for large Hubbard interaction $U/t$, but at the same time the spin diffusivity $D \gtrsim \hbar/m_s$ is bounded from below with the effective mass $m_s=mU/t$ of spin excitations \cite{Nichols:2018}.

Beyond conductors, an open question is the {\bf applicability of transport bounds to spin-insulating phases}, such as superfluids or ferromagnets. Spin-insulating behavior has been recently observed in the SF regime\cite{Krinner:2016,PhysRevLett.117.255302,Zhang:2017,Ueda:2017}, but local spin transport bounds have yet to be explored. Measurements of longitudinal spin diffusivity, conductivity, and susceptibility in the SF phase would test the wide range of predictions from theory \cite{Wulin:2011,Enss:2012lw,Wlazlowski:2013, Jensen:2018} as discussed in \S\ref{sec:bounds}. Also in the deeply degenerate regime, the low-temperature \textbf{anisotropy effect} discussed in \S\ref{sec:FLT} has not been observed in ultracold atoms, and would be useful to compare against its observation \cite{Akimoto:2003} in liquid helium, and test our understanding of the distinction between $D^\perp$ and $D^\parallel$. 

In a broader context, universal quantum bounds on transport are found in holographic quantum matter \cite{Hartnoll:2016}, which is characterized by the absence of quasiparticles.
Since quasi-particles are long-lived excitations, their absence is associated with fast relaxation times. 
Generically, the equilibration time can become as short as $\tau_\varphi \geq C \hbar/k_B T$ for $T\to0$, and systems without quasiparticles can saturate this bound.  
We have discussed how spin transport in strongly interacting Fermi gases saturate this bound.  
Similarly, in certain systems without quasiparticles the Lyapunov time $\tau_L$ to lose memory of the initial state saturates the bound $\tau_L \geq \frac1{2\pi} \hbar/k_B T$, hence such systems reach quantum chaos in the shortest possible time \cite{Hartnoll:2016}.  While conserved quantities relax hydrodynamically, all other generic perturbations decay quickly and locally on a timescale of $\tau_\varphi \gtrsim \hbar/k_B T$.  There are bounds on shear viscosity, electrical conductivity, and other transport coefficients which can be mapped to holographic duals; here we have reviewed quantitative results for bounds on spin diffusion in strongly interacting Fermi gases. Connecting these lines of inquiry raises the question of {\bf whether there exist holographic duals with SU(2) spin symmetry, and how they compare to experimental observations to date.}

We have discussed how bounds on a shortest possible microscopic scattering time $\tau$ manifest themselves as lower bounds on \emph{local} diffusivity and conductivity $\sigma \propto \tau$.  At the same time, fast local equilibration leads to a hydrodynamic evolution with a long relaxation time of \emph{global} collective motion, as in the spin-dipole damping $\Gamma_\text{SD}\simeq \omega_z^2\tau$ of a trapped gas with trapping frequency $\omega_z$ in \S\ref{sec:trap}. Similarly, the global bulk relaxation rate of transverse magnetization, proportional to $R_M = (D\alpha^2)^{1/3} \sim \tau^{1/3}$, is small for fast local equilibration.
In this case, the lower bound on microscopic scattering time leads to an \emph{upper} bound on the global relaxation time. In the case of heat conductance through a channel, this can be seen as a bound on the rate of information flow \cite{Taylor:2015}. A broader question raised is, {\bf Are there universal bounds on global relaxation rates in spin systems?} An ideal system in which to explore such questions is the two-terminal geometry, reviewed in \cite{Krinner:2017} and in another article of this volume.

Our discussion of transport coefficients in \S\ref{sec:TransCoeffs} often relied on the elegant scale invariance of the unitary Fermi gas. Extending to other systems, a key question is {\bf What is the role of scale invariance in transport bounds?} The viscosity bound was first found in supersymmetric Yang-Mills theories which are scale (conformal) invariant \cite{Schaefer:2009}, so the strongly interacting Fermi gas at the scale invariant unitary QCP is a natural candidate for transport bounds. Later, it was found that the diffusion minima lie not exactly at unitarity but slightly on the BEC side where medium scattering is even stronger \cite{Valtolina:2017}.  The transport minimum is thus facilitated by the vicinity of the scale invariant QCP, but the lowest numerical value is found at a small perturbation $\lambda_T/a>0$ away from it.

This raises the question whether transport bounds can be expected in the strongly interacting \emph{two-dimensional} Fermi gas.  The 2D quantum gas with zero-range attractive contact interaction is scale invariant at the classical level, but in quantum mechanics the scattering potential always admits a bound state with binding energy $\eb=\hbar^2/m\a2d^2>0$.  This new energy scale, and the 2D scattering length $a_\text{2D}$, break scale invariance and are a manifestation of a quantum scale anomaly.  The 2D Fermi gas has only two trivial fixed points, one for the very weakly attractive, almost ideal Fermi gas as $\a2d\to\infty$ and the other for the very strongly attractive Fermi gas, which becomes a weakly repulsive gas of bosonic molecules as $\a2d\to0$.  In particular, there is no interacting quantum critical point in the middle of the BCS-BEC crossover resembling the unitary Fermi gas in 3D.  It is a very interesting question whether transport bounds still exist without a scale-invariant QCP, and in fact the first experiment on transverse diffusion found a minimum $D_\perp \gtrsim 0.0063\hbar/m$ \cite{Koschorreck2013} roughly a hundred times below the expected bound. Still, the 2D Fermi gas has a strongly interacting regime when the three length scales $n^{-1/2} \sim \lambda_T \sim \a2d$ are all comparable \cite{Bauer:2014,Levinsen:2015dt}, and a later experiment found $D_0^\perp \gtrsim 1.7\hbar/m$  \cite{Luciuk:2017iw}, instead consistent with a generic quantum bound for the scattering rate $\tau_\perp^{-1} \lesssim k_B T/\hbar \sim \ef/\hbar \sim \eb/\hbar$ in a strongly interacting quantum fluid. With only two conflicting measurements, the open experimental question remains: {\bf Are 2D systems compatible with the \bm{$D \gtrsim \hbar/m$} bound?} Transverse diffusivity in 2D should be measured by other groups; and 2D longitudinal diffusivity has yet to be measured.

In summary, spin transport in unitary ultracold fermions provides a specific and controlled example of universality in strongly interacting non-equilibrium systems. The salient phenomenon of bounded transport coefficients connects to a broader discussion of saturated local equilibration times. Understanding the range of applicability of this paradigm and a quantitative understanding of the bounds are among the most important open questions.

\begin{summary}[SUMMARY POINTS]
\begin{enumerate}
\item Longitudinal and transverse spin dynamics are distinguished by the Leggett-Rice effect, that causes precession of the transverse spin current around the local magnetization, and by anisotropy at low temperate due to Fermi statistics.
\item Several experimental studies of spin dynamics support the existence of a lower bound on spin diffusivity, and upper bound on local spin drag, in normal-state fermions with resonant $s$-wave interactions.
\item The observed rates of global relaxation of magnetization are consistent with a lower bound on local relaxation time, a phenomenon observed in diverse systems including incoherent conductors, bad metals, heavy-fermion compounds, and high-energy plasmas.
\end{enumerate}
\end{summary}

\section*{DISCLOSURE STATEMENT}
The authors are not aware of any affiliations, memberships, funding, or financial holdings that might be perceived as affecting the objectivity of this review.
%
\section*{ACKNOWLEDGMENTS}
We thank the research groups at Aarhus, Bonn, LENS, and MIT for permission to use their data to make the figures in this review, and thank Eugene Demler, Arun Paramekanti, Edward Taylor, Shizhong Zhang, Wilhelm Zwerger, and Martin Zwierlein for stimulating conversations. This work is part of and supported by the DFG Collaborative Research Centre SFB 1225 (ISOQUANT), and supported by NSERC, by AFOSR under FA9550-13-1-0063, and by ARO under W911NF-15-1-0603.
\bibliography{STRev}

\begin{thebibliography}{120}
\expandafter\ifx\csname natexlab\endcsname\relax\def\natexlab#1{#1}\fi

\bibitem{Hahn:1950tv}
Hahn EL. 1950.
{Spin Echoes}.
\textit{Phys. Rev.} 80:580--594

\bibitem{Carr:1954wf}
Carr HY, Purcell EM. 1954.
Effects of diffusion on free precession in nuclear magnetic resonance
  experiments.
\textit{Phys. Rev.} 94:630--638

\bibitem{Torrey:1956tk}
Torrey HC. 1956.
Bloch equations with diffusion terms.
\textit{Phys. Rev.} 104:563--565

\bibitem{Silin:1958tc}
Silin VP. 1958.
{Oscillations of a Fermi liquid in a magnetic field}.
\textit{Sov. Phys. JETP} 6:945

\bibitem{Hone:1961fx}
Hone D. 1961.
{Self-Diffusion in Liquid $^3$He}.
\textit{Phys. Rev.} 121:669--673

\bibitem{LR:1968}
Leggett AJ, Rice MJ. 1968.
{Spin Echoes in Liquid $^3$He and Mixtures: A Predicted New Effect}.
\textit{Phys. Rev. Lett.} 20:586--589

\bibitem{Leggett:1970}
Leggett AJ. 1970.
{Spin diffusion and spin echoes in liquid $^3$He at low temperature}.
\textit{J. Phys. C} 3:448--459

\bibitem{Corruccini:1971}
Corruccini LR, Osheroff DD, Lee DM, Richardson RC. 1971.
{Spin Diffusion in Liquid $^3$He: The Effect of Leggett and Rice}.
\textit{Phys. Rev. Lett.} 27:650--653

\bibitem{Corruccini:1972vt}
Corruccini LR, Osheroff DD, Lee DM, Richardson RC. 1972.
{Spin-wave phenomena in liquid $^3$He systems}.
\textit{J. Low Temp. Phys.} 8:229--254

\bibitem{3HeWheatley}
Wheatley JC. 1975.
{Experimental properties of superfluid $^3$He}.
\textit{Rev. Mod. Phys.} 47:415--470

\bibitem{Meyerovich:1991}
Meyerovich AE. 1991.
{Perspectives of the theory of spin-polarized quantum systems: beyond the
  standard Boltzmann equation}.
\textit{Physica B} 169:183--189

\bibitem{OwersBradley:1999uu}
Owers-Bradley JR. 1999.
{Spin polarized liquids}.
\textit{Rep. Prog. Phys.} 60:1173

\bibitem{Bashkin:1981}
Bashkin EP. 1981.
Spin waves in polarized paramagnetic gases.
\textit{JETP Lett.} 33:11

\bibitem{Laloe:1982ww}
Lhuillier C, Lalo\"{e} F. 1982{\natexlab{a}}.
{Transport properties in a spin polarized gas, I}.
\textit{J. Phys. (Paris)} 43:197--224

\bibitem{Laloe:1982wc}
Lhuillier C, Lalo\"{e} F. 1982{\natexlab{b}}.
{Transport properties in a spin-polarized gas, II}.
\textit{J. Phys. (Paris)} 43:225--241

\bibitem{Johnson:1984gr}
Johnson BR, Denker JS, Bigelow N, L{\'e}vy LP, Freed JH, Lee DM. 1984.
{Observation of Nuclear Spin Waves in Spin-Polarized Atomic Hydrogen Gas}.
\textit{Phys. Rev. Lett.} 52:1508--1511

\bibitem{Levy:1984du}
L{\'e}vy LP, Ruckenstein AE. 1984.
{Collective Spin Oscillations in Spin-Polarized Gases: Spin-Polarized
  Hydrogen}.
\textit{Phys. Rev. Lett.} 52:1512

\bibitem{Meyerovich:1985}
Meyerovich AE. 1985.
{Degeneracy effects in the spin dynamics of spin-polarized Fermi gases}.
\textit{Phys. Lett. A} 107:177--180

\bibitem{Miyake:1985}
Miyake K, Mullin WJ, Stamp PCE. 1985.
{Mean-field and spin-rotation phenomena in Fermi systems: the relation between
  the Leggett-Rice and Lhuillier-Lalo{\"e} effects}.
\textit{J. Phys. (Paris)} 46:663--671

\bibitem{Vasiliev:2012}
Vainio O, Ahokas J, Novotny S, Sheludyakov S, Zvezdov D, et~al. 2012.
{Guiding and Trapping of Electron Spin Waves in Atomic Hydrogen Gas}.
\textit{Phys. Rev. Lett.} 108:185304

\bibitem{Mcguirk:2002}
McGuirk JM, Lewandowski HJ, Harber DM, Nikuni T, Williams JE, Cornell EA. 2002.
Spatial resolution of spin waves in an ultracold gas.
\textit{Phys. Rev. Lett.} 89:090402

\bibitem{Du:2008de}
Du X, Luo L, Clancy B, Thomas JE. 2008.
{Observation of anomalous spin segregation in a trapped {F}ermi gas}.
\textit{Phys. Rev. Lett.} 101:150401

\bibitem{Hulet2011}
Liao YA, Revelle M, Paprotta T, Rittner ASC, Li W, et~al. 2011.
Metastability in spin-polarized {F}ermi gases.
\textit{Phys. Rev. Lett.} 107:145305

\bibitem{Koschorreck2013}
Koschorreck M, Pertot D, Vogt E, K\"{o}hl M. 2013.
{Universal spin dynamics in two-dimensional {F}ermi gases}.
\textit{Nature Phys.} 9:405--409

\bibitem{Sommer:2011}
Sommer A, Ku M, Roati G, Zwierlein MW. 2011.
{Universal spin transport in a strongly interacting Fermi gas}.
\textit{Nature} 472:201--204

\bibitem{Sommer:2011njp}
Sommer A, Ku M, Zwierlein MW. 2011.
{Spin transport in polaronic and superfluid Fermi gases}.
\textit{New J. Phys.} 13:055009

\bibitem{Bardon:2014}
Bardon AB, Beattie S, Luciuk C, Cairncross W, Fine D, et~al. 2014.
{Transverse demagnetization dynamics of a unitary Fermi gas}.
\textit{Science} 344:722

\bibitem{Trotzky:2015fe}
Trotzky S, Beattie S, Luciuk C, Smale S, Bardon AB, et~al. 2015.
{Observation of the Leggett-Rice Effect in a Unitary Fermi Gas}.
\textit{Phys. Rev. Lett.} 114:015301

\bibitem{Luciuk:2017iw}
Luciuk C, Smale S, B{\"o}ttcher F, Sharum H, Olsen BA, et~al. 2017.
{Observation of Quantum-Limited Spin Transport in Strongly Interacting
  Two-Dimensional Fermi Gases}.
\textit{Phys. Rev. Lett.} 118:130405

\bibitem{Du:2009hm}
Du X, Zhang Y, Petricka J, Thomas JE. 2009.
{Controlling spin current in a trapped {F}ermi gas}.
\textit{Phys. Rev. Lett.} 103:010401

\bibitem{Rosenbusch:2010}
Deutsch C, Ramirez-Martinez F, Lacro\^ute C, Reinhard F, Schneider T, et~al.
  2010.
Spin self-rephasing and very long coherence times in a trapped atomic ensemble.
\textit{Phys. Rev. Lett.} 105:020401

\bibitem{Krauser:2014ca}
Krauser JS, Ebling U, Fl{\"a}schner N, Heinze J, Sengstock K, et~al. 2014.
{Giant spin oscillations in an ultracold Fermi sea}.
\textit{Science} 343:157

\bibitem{Taylor:2015}
Taylor E, Segal D. 2015.
{Quantum bounds on heat transport through nanojunctions}.
\textit{Phys. Rev. Lett.} 114:220401

\bibitem{Krinner:2017}
Krinner S, Esslinger T, Brantut JP. 2017.
Two-terminal transport measurements with cold atoms.
\textit{J. Phys.: Condens. Matter} 29:343003

\bibitem{LewensteinLatticeReview}
Lewenstein M, Sanpera A, Ahufinger V, Damski B, Sen(De) A, Sen U. 2007.
Ultracold atomic gases in optical lattices: mimicking condensed matter physics
  and beyond.
\textit{Advances in Physics} 56:243--379

\bibitem{PolkovnikovRev}
Polkovnikov A, Sengupta K, Silva A, Vengalattore M. 2011.
{Colloquium: Nonequilibrium dynamics of closed interacting quantum systems}.
\textit{Rev. Mod. Phys.} 83:863--883

\bibitem{spintronics}
van~der Straten P. 2013.
{Condensed-matter physics: Spintronics, the atomic way}.
\textit{Nature} 498:175

\bibitem{kondo}
Foss-Feig M, Hermele M, Rey AM. 2010.
{Probing the Kondo lattice model with alkaline-earth-metal atoms}.
\textit{Phys. Rev. A} 81:051603

\bibitem{Bruun:2008}
Bruun GM, Recati A, Pethick CJ, Smith H, Stringari S. 2008.
{Collisional properties of a polarized Fermi gas with resonant interactions}.
\textit{Phys. Rev. Lett.} 100:240406

\bibitem{Oktel:2002}
Oktel M{\"O}, Levitov LS. 2002.
Internal waves and synchronized precession in a cold vapor.
\textit{Phys. Rev. Lett.} 88:230403

\bibitem{Fuchs:2002}
Fuchs JN, Gangardt DM, Lalo{\"e} F. 2002.
Internal state conversion in ultracold gases.
\textit{Phys. Rev. Lett.} 88:230404

\bibitem{Williams:2002}
Williams JE, Nikuni T, Clark CW. 2002.
{Longitudinal spin waves in a dilute Bose gas}.
\textit{Phys. Rev. Lett.} 88:230405

\bibitem{Fuchs:2003}
Fuchs JN, Gangardt DM, Lalo{\"e} F. 2003.
{Large amplitude spin waves in ultra-cold gases}.
\textit{Eur. Phys. J. D} 25:57--75

\bibitem{Piechon:2009}
Pi{\'e}chon F, Fuchs JN, Lalo{\"e} F. 2009.
{Cumulative Identical Spin Rotation Effects in Collisionless Trapped Atomic
  Gases}.
\textit{Phys. Rev. Lett.} 102:215301

\bibitem{Natu:2009}
Natu SS, Mueller EJ. 2009.
{Anomalous spin segregation in a weakly interacting two-component Fermi gas}.
\textit{Phys. Rev. A} 79:051601

\bibitem{Heinze:2013ko}
Heinze J, Krauser JS, Flaschner N, Sengstock K, Becker C, et~al. 2013.
{Engineering Spin Waves in a High-Spin Ultracold Fermi Gas}.
\textit{Phys. Rev. Lett.} 110:250402

\bibitem{Ebling:2014}
Ebling U, Krauser JS, Fl{\"a}schner N, Sengstock K, Becker C, et~al. 2014.
{Relaxation dynamics of an isolated large-spin Fermi gas far from equilibrium}.
\textit{Phys. Rev. X} 4:021011

\bibitem{Niroomand:2015fu}
Niroomand D, Graham SD, McGuirk JM. 2015.
{Longitudinal Spin Diffusion in a Nondegenerate Trapped $^{87}$Rb Gas}.
\textit{Phys. Rev. Lett.} 115:075302

\bibitem{Xu:2015}
Xu J, Gu Q, Mueller EJ. 2015.
Collisionless spin dynamics in a magnetic field gradient.
\textit{Phys. Rev. A} 91:043613

\bibitem{Koller:2015}
Koller AP, Mundinger J, Wall ML, Rey AM. 2015.
Demagnetization dynamics of noninteracting trapped fermions.
\textit{Phys. Rev. A} 92:033608

\bibitem{Xu:2017}
Xu J, Gu Q. 2017.
Crossover from collisionless to collisional spin dynamics of polarized
  fermions.
\textit{Europhys. Lett.} 117:10008

\bibitem{Graham:2018}
Graham SD, Niroomand D, Ragan RJ, McGuirk JM. 2018.
Stable spin domains in a non-degenerate ultra-cold gas.
\textit{arXiv:1803.01096}

\bibitem{Fomin:1997}
Fomin IA. 1997.
{Transverse spin dynamics of a spin-polarized Fermi liquid}.
\textit{JETP Lett.} 65:749

\bibitem{Mullin:2005}
Mullin WJ, Ragan RJ. 2005.
Derivation of transverse spin-wave dynamics from a kinetic equation in a
  rotating reference frame.
\textit{J. Low Temp. Phys.} 138:73--78

\bibitem{BruunPeth:2011}
Bruun GM, Pethick CJ. 2011.
{Spin Diffusion in Trapped Clouds of Cold Atoms with Resonant Interactions}.
\textit{Phys. Rev. Lett.} 107:255302

\bibitem{Valtolina:2017}
Valtolina G, Scazza F, Amico A, Burchianti A, Recati A, et~al. 2017.
{Exploring the ferromagnetic behaviour of a repulsive Fermi gas through spin
  dynamics}.
\textit{Nature Phys.} 13:704

\bibitem{Vichi:1999}
Vichi L, Stringari S. 1999.
{Collective oscillations of an interacting trapped Fermi gas}.
\textit{Phys. Rev. A} 60:4734

\bibitem{Heiselberg:2012}
Heiselberg H. 2012.
Inhomogeneous spin diffusion in traps with cold atoms.
\textit{Phys. Rev. Lett.} 108:245303

\bibitem{Enss:2015er}
Enss T. 2015.
{Nonlinear spin diffusion and spin rotation in a trapped Fermi gas}.
\textit{Phys. Rev. A} 91:023614

\bibitem{Recati:2011do}
Recati A, Stringari S. 2011.
{Spin Fluctuations, Susceptibility, and the Dipole Oscillation of a Nearly
  Ferromagnetic Fermi Gas}.
\textit{Phys. Rev. Lett.} 106:080402

\bibitem{Christensen:2015}
Christensen RS, Bruun GM. 2015.
{Quasiparticle scattering rate in a strongly polarized Fermi mixture}.
\textit{Phys. Rev. A} 91:042702

\bibitem{Taylor:2011}
Taylor E, Zhang S, Schneider W, Randeria M. 2011.
Colliding clouds of strongly interacting spin-polarized fermions.
\textit{Phys. Rev. A} 84:063622

\bibitem{Goulko:2011}
Goulko O, Chevy F, Lobo C. 2011.
{Collision of two spin-polarized fermionic clouds}.
\textit{Phys. Rev. A} 84:051605

\bibitem{Goulko:2013}
Goulko O, Chevy F, Lobo C. 2013.
{Spin drag of a Fermi gas in a harmonic trap}.
\textit{Phys. Rev. Lett.} 111:190402

\bibitem{Schaefer:2016ga}
Schaefer T. 2016.
{Generalized theory of diffusion based on kinetic theory}.
\textit{Phys. Rev. A} 94:043644

\bibitem{Bamler:2015}
Bamler R, Rosch A. 2015.
{Equilibration and approximate conservation laws: Dipole oscillations and
  perfect drag of ultracold atoms in a harmonic trap}.
\textit{Phys. Rev. A} 91:063604

\bibitem{Enss:2012lw}
Enss T, Haussmann R. 2012.
{Quantum Mechanical Limitations to Spin Transport in the Unitary Fermi Gas}.
\textit{Phys. Rev. Lett.} 109:195303

\bibitem{Sachdev:1999}
Sachdev S. 1999.
{Quantum Phase Transitions}.
Cambridge: Cambridge University Press

\bibitem{Zwerger:2016}
Zwerger W. 2016.
{Strongly Interacting Fermi Gases}. In \textit{{Quantum Matter at Ultralow
  Temperatures}}, eds. M~Inguscio, W~Ketterle, S~Stringari, G~Roati, vol. 191
  of \textit{Proceedings of the International School of Physics ``Enrico
  Fermi'' Course 191}. Amsterdam: IOS Press

\bibitem{Gezerlis:2014}
Gezerlis A, Pethick CJ, Schwenk A. 2014.
Pairing and superfluidity of nucleons in neutron stars.
\textit{arXiv:1406.6109}

\bibitem{Strinati:2018}
Strinati GC, Pieri P, R{\"o}pke G, Schuck P, Urban M. 2018.
{The BCS-BEC crossover: From ultra-cold Fermi gases to nuclear systems}.
\textit{Physics Reports} 738:1--76

\bibitem{Baym:2008}
Baym G, Pethick C. 2008.
{Landau Fermi-Liquid Theory: Concepts and Applications}.
New York: John Wiley \& Sons

\bibitem{Jeon:1989}
Jeon JW, Mullin WJ. 1989.
{Transverse spin diffusion in polarized Fermi gases}.
\textit{Phys. Rev. Lett.} 62:2691--2694

\bibitem{Mullin:2006}
Mullin WJ, Ragan RJ. 2006.
{Spin diffusion in trapped gases: Anisotropy in dipole and quadrupole modes}.
\textit{Phys. Rev. A} 74:043607

\bibitem{Bruun:2011}
Bruun GM. 2011.
{Spin diffusion in Fermi gases}.
\textit{New J. Phys.} 13:035005

\bibitem{Enss:2013}
Enss T. 2013{\natexlab{a}}.
{Transverse spin diffusion in strongly interacting Fermi gases}.
\textit{Phys. Rev. A} 88:033630

\bibitem{Bruun:2012}
Bruun GM. 2012.
{Shear viscosity and spin diffusion coefficient of a two-dimensional Fermi
  gas}.
\textit{Phys. Rev. A} 85:013636

\bibitem{Enss:2012a}
Enss T, K\"uppersbusch C, Fritz L. 2012.
Shear viscosity and spin diffusion in a two-dimensional {F}ermi gas.
\textit{Phys. Rev. A} 86:013617

\bibitem{Parish:2009}
Parish MM, Huse DA. 2009.
{Evaporative depolarization and spin transport in a unitary trapped Fermi gas}.
\textit{Phys. Rev. A} 80:063605

\bibitem{Fu:1974}
Fu HH, Ebner C. 1974.
{Transport coefficients in a two-dimensional Fermi liquid}.
\textit{Phys. Rev. A} 10:338

\bibitem{Miyake:1984}
Miyake K, Mullin WJ. 1984.
{Theory of a polarized Fermi liquid in two dimensions: Spin diffusion}.
\textit{J. Low Temp. Phys.} 56:499--522

\bibitem{Meyerovich:1994}
Meyerovich AE, Musaelian KA. 1994.
{Anomalous spin dynamics and relaxation in Fermi liquids}.
\textit{Phys. Rev. Lett.} 72:1710

\bibitem{Forster:1975}
Forster D. 1975.
{Hydrodynamic fluctuations, broken symmetry, and correlation functions}.
Reading: WA Benjamin

\bibitem{Mineev:2004}
Mineev VP. 2004.
{Transverse spin dynamics in a spin-polarized Fermi liquid}.
\textit{Phys. Rev. B} 69:144429

\bibitem{Akimoto:2003}
Akimoto H, Candela D, Xia JS, Mullin WJ, Adams ED, Sullivan NS. 2003.
{New evidence for zero-temperature relaxation in a spin-polarized Fermi
  liquid}.
\textit{Phys. Rev. Lett.} 90:105301

\bibitem{Krinner:2016}
Krinner S, Lebrat M, Husmann D, Grenier C, Brantut JP, Esslinger T. 2016.
Mapping out spin and particle conductances in a quantum point contact.
\textit{Proc. Nat. Acad. Sci.} 113:8144--8149

\bibitem{PhysRevLett.117.255302}
Kan\'asz-Nagy M, Glazman L, Esslinger T, Demler EA. 2016.
{Anomalous Conductances in an Ultracold Quantum Wire}.
\textit{Phys. Rev. Lett.} 117:255302

\bibitem{Zhang:2017}
Liu B, Zhai H, Zhang S. 2017.
{Anomalous conductance of a strongly interacting Fermi gas through a quantum
  point contact}.
\textit{Phys. Rev. A} 95:013623

\bibitem{Ueda:2017}
Uchino S, Ueda M. 2017.
{Anomalous Transport in the Superfluid Fluctuation Regime}.
\textit{Phys. Rev. Lett.} 118:105303

\bibitem{Einzel:1991}
Einzel D. 1991.
{The spin diffusion in normal and superfluid Fermi liquids}.
\textit{J. Low Temp. Phys.} 84:321--356

\bibitem{Wulin:2011}
Wulin D, Guo H, Chien CC, Levin K. 2011.
{Spin transport in cold Fermi gases: A pseudogap interpretation of spin
  diffusion experiments at unitarity}.
\textit{Phys. Rev. A} 83:061601

\bibitem{Mueller:pseudogaps}
Mueller EJ. 2017.
{Review of pseudogaps in strongly interacting Fermi gases}.
\textit{Rep. Prog. Phys.} 80:104401

\bibitem{Duine:2010}
Duine RA, Polini M, Stoof HTC, Vignale G. 2010.
{Spin drag in an ultracold Fermi gas on the verge of ferromagnetic
  instability}.
\textit{Phys. Rev. Lett.} 104:220403

\bibitem{Mineev:2005js}
Mineev V. 2005.
{Theory of transverse spin dynamics in a polarized Fermi liquid and an
  itinerant ferromagnet}.
\textit{Phys. Rev. B} 72:144418

\bibitem{Nikolic:2007}
Nikoli{\'c} P, Sachdev S. 2007.
{Renormalization-group fixed points, universal phase diagram, and $1/N$
  expansion for quantum liquids with interactions near the unitarity limit}.
\textit{Phys. Rev. A} 75:033608

\bibitem{Ku:2012}
Ku MJH, Sommer AT, Cheuk LW, Zwierlein MW. 2012.
{Revealing the Superfluid Lambda Transition in the Universal Thermodynamics of
  a Unitary Fermi Gas}.
\textit{Science} 335:563--567

\bibitem{Enss:2012crit}
Enss T. 2012.
{Quantum critical transport in the unitary Fermi gas}.
\textit{Phys. Rev. A} 86:013616

\bibitem{Zaanen:2004he}
Zaanen J. 2004.
{Superconductivity - Why the temperature is high}.
\textit{Nature} 430:512

\bibitem{Farinas:1999}
Farinas PF, Bedell KS, Studart N. 1999.
{Hidden spin-current conservation in 2d Fermi liquids}.
\textit{Phys. Rev. Lett.} 82:3851

\bibitem{Hofmann:2011}
Hofmann J. 2011.
{Current response, structure factor and hydrodynamic quantities of a two- and
  three-dimensional Fermi gas from the operator-product expansion}.
\textit{Phys. Rev. A} 84:043603

\bibitem{Enss:2013sumrule}
Enss T. 2013{\natexlab{b}}.
{Shear viscosity and spin sum rules in strongly interacting Fermi gases}.
\textit{Eur. Phys. J. Special Topics} 217:169--175

\bibitem{Wlazlowski:2013}
Wlaz{\l}owski G, Magierski P, Drut JE, Bulgac A, Roche KJ. 2013.
{Cooper pairing above the critical temperature in a unitary Fermi gas}.
\textit{Phys. Rev. Lett.} 110:090401

\bibitem{Mink:2012}
Mink MP, Jacobs VPJ, Stoof HTC, Duine RA, Polini M, Vignale G. 2012.
{Spin transport in a unitary Fermi gas close to the BCS transition}.
\textit{Phys. Rev. A} 86:063631

\bibitem{Mink:2013}
Mink MP, Stoof HTC, Duine RA, Polini M, Vignale G. 2013.
{Unified Boltzmann transport theory for the drag resistivity close to an
  interlayer-interaction-driven second-order phase transition}.
\textit{Phys. Rev. B} 88:235311

\bibitem{Schaefer:2009}
Sch{\"a}fer T, Teaney D. 2009.
{Nearly perfect fluidity: from cold atomic gases to hot quark gluon plasmas}.
\textit{Rep. Prog. Phys.} 72:126001

\bibitem{Enss:2011}
Enss T, Haussmann R, Zwerger W. 2011.
{Viscosity and scale invariance in the unitary Fermi gas}.
\textit{Ann. Phys. (NY)} 326:770--796

\bibitem{Jensen:2018}
Jensen S, Gilbreth CN, Alhassid Y. 2018.
{Nature of pairing correlations in the homogeneous Fermi gas at unitarity}.
\textit{arXiv:1801.06163}

\bibitem{Bruin:2013}
Bruin JAN, Sakai H, Perry RS, Mackenzie AP. 2013.
{Similarity of scattering rates in metals showing T-linear resistivity}.
\textit{Science} 339:804--807

\bibitem{Hartnoll:2015}
Hartnoll SA. 2015.
Theory of universal incoherent metallic transport.
\textit{Nature Phys.} 11:54

\bibitem{Hartman:2017}
Hartman T, Hartnoll SA, Mahajan R. 2017.
Upper bound on diffusivity.
\textit{Phys. Rev. Lett.} 119:141601

\bibitem{Fukuhara:2013}
Fukuhara T, Kantian A, Endres M, Cheneau M, Schau{\ss} P, et~al. 2013.
Quantum dynamics of a mobile spin impurity.
\textit{Nature Phys.} 9:235

\bibitem{Hild:2014it}
Hild S, Fukuhara T, Schau{\ss} P, Zeiher J, Knap M, et~al. 2014.
{Far-from-Equilibrium Spin Transport in Heisenberg Quantum Magnets}.
\textit{Phys. Rev. Lett.} 113:147205

\bibitem{Nichols:2018}
Nichols MA, Cheuk LW, Okan M, Hartke TR, Mendez E, et~al. 2018.
{Spin Transport in a Mott Insulator of Ultracold Fermions}.
\textit{arXiv:1802.10018}

\bibitem{anderson2017optical}
Anderson R, Wang F, Xu P, Venu V, Trotzky S, et~al. 2017.
Optical conductivity of a quantum gas.
\textit{arXiv:1712.09965}

\bibitem{brown2018bad}
Brown PT, Mitra D, Guardado-Sanchez E, Nourafkan R, Reymbaut A, et~al. 2018.
{Bad metallic transport in a cold atom Fermi-Hubbard system}.
\textit{arXiv:1802.09456}

\bibitem{Emery:1995prl}
Emery VJ, Kivelson SA. 1995.
Superconductivity in bad metals.
\textit{Phys. Rev. Lett.} 74:3253

\bibitem{Perepelitsky:2016}
Perepelitsky E, Galatas A, Mravlje J, Khatami E, {Sriram Shastry} B, Georges A.
  2016.
{Transport and optical conductivity in the Hubbard model: A high-temperature
  expansion perspective}.
\textit{Phys. Rev. B} 94:235115

\bibitem{Hartnoll:2016}
Hartnoll SA, Lucas A, Sachdev S. 2016.
Holographic quantum matter.
\textit{arXiv:1612.07324}

\bibitem{Bauer:2014}
Bauer M, Parish MM, Enss T. 2014.
{Universal Equation of State and Pseudogap in the Two-Dimensional Fermi Gas}.
\textit{Phys. Rev. Lett.} 112:135302

\bibitem{Levinsen:2015dt}
Levinsen J, Parish MM, Parish M. 2015.
{Strongly interacting two-dimensional Fermi gases}.
\textit{Ann. Rev. Cold Atoms Mols.} 3:1--75

\end{thebibliography}

\end{document}